\documentclass{iopart} 
\usepackage{iopams}
\usepackage{graphicx}
\usepackage{color}
\begin{document}

\topical[Laser-MBE of ZnO thin films and heterostructures]{Laser molecular beam epitaxy of ZnO thin films and heterostructures }

\author{Matthias Opel$^1$,
        Stephan Gepr\"{a}gs$^1$,
        Matthias Althammer$^1$\footnote{Present address: University of Alabama, Center for Materials for Information Technology MINT, Tuscaloosa, AL 35487 USA},
        Thomas Brenninger$^1$
        and Rudolf Gross$^{1,2}$}

\address{$^1$
         Walther-Mei{\ss}ner-Institut,
         Bayerische Akademie der Wissenschaften,
         85748 Garching,
         Germany}
\address{$^2$
         Physik-Department,
         Technische Universit\"{a}t M\"{u}nchen,
         85748 Garching,
         Germany}

\ead{Matthias.Opel@wmi.badw.de, Rudolf.Gross@wmi.badw.de}

\begin{abstract} 
We report on the growth of epitaxial ZnO thin films and ZnO based heterostructures on sapphire substrates by laser molecular beam epitaxy (MBE). We first discuss some recent developments in laser-MBE such as flexible ultra-violet laser beam optics, infrared laser heating systems or the use of atomic oxygen and nitrogen sources, and describe the technical realization of our advanced laser-MBE system. Then we describe the optimization of the deposition parameters for ZnO films such as laser fluence and substrate temperature and the use of buffer layers. The detailed structural characterization by x-ray analysis and transmission electron microscopy shows that epitaxial ZnO thin films with high structural quality can be achieved, as demonstrated by a small out-of-plane and in-plane mosaic spread as well as the absence of rotational domains. We also demonstrate the heteroepitaxial growth of ZnO based multilayers as a prerequisite for spin transport experiments and the realization of spintronic devices. As an example, we show that TiN/Co/ZnO/Ni/Au multilayer stacks can be grown on (0001)-oriented sapphire with good structural quality of all layers and well defined in-plane epitaxial relations.
\vspace*{5mm}\\
Date: \today.
\end{abstract}

\vspace{2pc}
\submitto{\JPD}
\maketitle


\section{Introduction} \label{sec:intro}

The direct band gap II-VI semiconductor zinc oxide (ZnO) is a versatile material for device applications \cite{Look2001,Pearton2004,Ozgur2005,Klingshirn2007,Janotti2009}. Within the last decade, the research activities on the wide bandgap semiconductor ZnO experienced a worldwide revival \cite{Janotti2009,Klingshirn2010}. One the one hand, there has been growing interest in the magnetic properties of transition metal doped ZnO and the self-organized growth in the form of nano rods \cite{Ozgur2005,Willander:2009a}. On the other hand, there was renewed interest in device physics and (opto/spin)-electronic applications \cite{Pearton2004,Klingshirn2007}. In particular, the availability of high-quality large bulk single crystals \cite{Park1967}, the strong luminescence demonstrated in optically pumped lasers \cite{Reynolds1996}, the proposal of realizing a room-temperature dilute magnetic semiconductor (DMS) by transition metal doping of ZnO \cite{Dietl2000}, and the prospects of gaining control over its electrical conductivity \cite{Tsukazaki2005} have brought ZnO back into the focus of current interest. However, neither room-temperature ferromagnetism nor $p$-type conductivity have been reproducibly realized so far because of the formation of secondary phases \cite{Opel2008,Ney2010} or the lack of stability of acceptor levels in the natively $n$-type ZnO material \cite{Ozgur2005,Janotti2009}, respectively. It has initially been suggested that the unintentional $n$-type doping is caused by the presence of oxygen vacancies or Zn interstitials \cite{Harrison1954,Hutson1957}. Recent studies, however, indicate that unintentional impurities, such as interstitial and substitutional hydrogen \cite{VanDeWalle2000}, or substitutional Al, Ga, and In \cite{McCluskey2007} act as shallow donors and might explain the $n$-type conduction in ZnO.

ZnO is widely used as a transparent conducting oxide \cite{Minami2005}. It shows a direct and wide band gap of $E_g=(3.365\pm 0.005)$\,eV at 300\,K in the near-ultraviolet range together with an electron mobility of 200\,cm$^2$/Vs and a large free-exciton binding energy of $(59.5\pm 0.5)$\,meV \cite{Thomas1960,Mang1995,Srikant1998,Reynolds1999}. Hence, excitonic emission processes can persist at or even above room temperature \cite{Reynolds1996}. Furthermore, a high electron mobility of 180,000\,cm$^2$/Vs was reported in (Mg,Zn)O/ZnO heterostructures together with the observation of a fractional quantum Hall effect \cite{Tsukazaki2010}. ZnO also displays a small spin-orbit coupling \cite{Fu2008} resulting in a large spin coherence length which is a prerequisite for the creation, transport and detection of spin-polarized currents in semiconductor spintronics \cite{Althammer2012}.

A key prerequisite for many important developments in ZnO research has been the improvement of material quality both in form of bulk single crystals and epitaxial thin films. Regarding the growth of oxide thin films and heterostructures, the tremendous progress achieved over the past decades is closely linked to the invention and further development of the appropriate deposition techniques \cite{Chen2013}. Because of their high melting points well above $1000^\circ$C, thermal evaporation of oxides is challenging. To overcome this problem, several alternatives (such as sputtering, spin coating, sol-gel processes, metal-organic chemical vapour deposition, molecular beam epitaxy) have been used, aiming at the controlled epitaxial growth of oxide thin films and heterostructures. To this end, an important impact originated from the development of pulsed laser deposition (PLD). Using a ruby laser, this technique has been applied to the growth of semiconductors and dielectrics already in 1965 \cite{Smith:1965a} and received broad attention after its successful application to the \textit{in-situ} growth of epitaxial high-temperature superconductor films in 1987 \cite{Dijkkamp:1987a}. Over the years, this technique has been refined by extending it to the ultra-high vacuum (UHV) regime and adding \textit{in-situ} analysis of the growth process. By this it was further developed into what today is called laser molecular beam epitaxy (laser-MBE) \cite{Gross2000}, allowing for the controlled layer-by-layer growth of oxide materials. Meanwhile, laser-MBE is one of the most promising techniques for the formation of complex-oxide heterostructures, superlattices, and well-controlled interfaces. Detailed information on its history and evolution is given in a variety of books \cite{Chrisey1994,Eason2006} and review articles \cite{Gross2000,Willmott:2000a,Christen2008,Martin2010,Opel2012}. Laser-MBE is based on a number of technical developments, including ultra-high vacuum (UHV) systems, pulsed ultraviolet (UV) lasers \cite{Basting:2001a,Dehmdahl:2008a}, high pressure RHEED (reflection high-energy electron diffraction) systems \cite{Rijnders:1997,Klein1999,Klein2000}, or infrared (IR) laser heating systems. Today it is successfully applied for the deposition of thin films and complex heterostructures of different classes of materials, including superconductors, various magnetic materials and ferroelectrics, on a large variety of appropriate single crystalline substrates.

The growth of ZnO thin films by laser-MBE turned out to be very successful because of the intrinsic flexibility of this deposition technique. In particular, transition-metal doping in ZnO thin films could be easily realized by growing from polycrystalline targets of different stoichiometry. Moreover, possible $p$-type conductivity could be investigated in ZnO films deposited in different background atmospheres or from different target materials, as well as by adding atomic gaseous species during deposition. An essential advantage hereby is the compatibility of the laser-MBE process with oxygen and other reactive gases.

\begin{figure}[tb]
    \centering{\includegraphics[width=6cm]{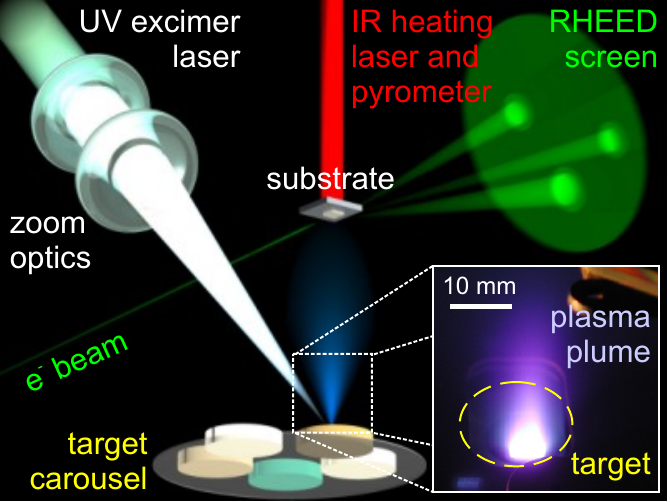}}
    \caption{\label{fig:pld}
             Schematic view inside the PLD process chamber operated at Walther-Mei{\ss}ner-Institut (WMI). The inset shows a photograph of the PLD plasma plume (reproduced from \cite{Opel2012}).}
\end{figure}

The typical arrangement of the main components inside a laser-MBE process chamber is illustrated in Figure~\ref{fig:pld}. The technique of PLD is conceptually simple and based on the stoichiometric transfer of material from a polycrystalline ``target'' to the surface of a substrate via the so-called ``plasma plume'' (inset in Figure~\ref{fig:pld}). Usually, several targets are placed on a rotatable plate (referred to as ``target carousel''), facilitating a quick change between different materials without changing the optical path for the UV laser beam. This is a particular advantage of PLD, allowing for the deposition of multiple oxide materials in a cost efficient way. During the ablation process, the laser beam can be scanned across the target while the target is rotated to allow for a homogeneous ablation of the target material. The substrate temperature is controlled via the substrate heater and determined by a pyrometer. Recently, contact-less infrared (IR) optical heating systems came into operation reaching temperatures of more than $1000^\circ$C. The relevant time scales are ns for the absorption of the UV laser light and the creation of the plasma (see below), $\mu$s for the material transfer to the substrate and typically ms for the diffusion processes on the substrate or the thin film surface.

Although PLD is conceptually simple, the various processes such as ablation, plasma formation, plume propagation, as well as nucleation and growth are complex. On irradiating the target material with a focused UV laser pulse, the photons are converted into electronic excitations rapidly relaxing into lattice, chemical, and mechanical energy \cite{Miotello:2010a}. This causes an abrupt ablation of the target material rather than an evaporation by heating. With laser fluences of several J/cm$^2$ and a laser pulse length in the 10\,ns regime, a huge laser power density of the order of $10^8$\,W/cm$^2$ is obtained. Taking a typical specific heat of 1\,J/gK and an absorption depth in the 10\,nm regime, this results in huge heating rates up to more than $10^{12}$\,K/s and instantaneous gas pressures typically ranging between 10 and 1000\,bar in the ablation process. This rough estimate clearly shows that the target material is not thermally evaporated but ablated from the target surface before reaching thermal equilibrium. This is a key advantage of PLD, because a stoichiometric transfer of the target material into the plasma plume and finally to the substrate surface is provided. We note that the laser-solid interaction mechanisms depend on the laser wavelength, determining the absorption depth. In order to keep the absorption depth small and to avoid subsurface heating, a short wavelength in the UV regime is used (e.g. provided by KrF (248\,nm) and ArF (193\,nm) excimer lasers). A further reduction of the wavelength is impractical since the absorption of the photons by the oxygen background atmosphere or optical elements in the beam path is increasing. We also note that for a long laser pulse duration (e.g. $\simeq 25$\,ns for a KrF excimer laser in contrast to only $\simeq 5$\,ns for a frequency-quadrupled Nd:YAG laser) there is a significant interaction of the incident laser pulse with the already ablated material, leading to a further heating of the plasma plume. The propagation of the plume can be studied by optical spectroscopy \cite{Geohegan:1996a}. It is found that neutral atoms, ions, and electrons travel at different velocities. They strongly interact with the background gas during propagation, resulting in thermalization. The thermalization of the high energetic plasma species is important to avoid resputtering of the growing film and strong intermixing at interfaces \cite{Fahler:1999a,Hau:1995a}.

The stoichiometric ablation of material from a solid target is a key factor for the success of PLD, allowing to grow complex oxide thin films \cite{Majewski2005,Geprags2009,Czeschka2009,Opel2011,Althammer2013}. For sufficiently high laser fluences resulting in a dense plasma and after a suitable pre-ablation of the target establishing a steady-state situation, the removal of target material is indeed preserving stoichiometry. However, this does not necessarily translate into the growth of a stoichiometric film. Since the sticking coefficients of the individual species may be different and since some re-sputtering may occur, not all elements may be incorporated into the film at the same rate. In particular, volatile elements (e.g.~Pb, Bi) may re-evaporate from the growing film surface. This can be compensated by using targets with an excess of the volatile component \cite{Geprags2007}. For oxide materials, the proper control of the oxygen content is most relevant. To this end, a key advantage of PLD is the fact that it tolerates a wide range of background pressures. Therefore, for some materials, PLD allows for better film properties than any other deposition method. Apart from type and pressure of the background atmosphere, important process parameters are the substrate temperature, the laser fluence at the target, the laser repetition rate, and the time intervals of growth and waiting (relaxation) cycles. A specific drawback of PLD is the generation of macroparticles, resulting from the explosive ejection of particles from the target, splashing and fragmentation due to thermal shock. In this context, the formation of macroscopic ``droplets'' with diameters in the micrometer range on top of the growing thin film is an omnipresent problem. Below we show that droplet formation can be avoided by optimizing the targets, by fine-tuning the process parameters and by carefully optimizing the laser spot on the target surface.

The use of ultra-high vacuum systems and the invention of advanced \textit{in situ} monitoring tools such as reflection high-energy electron diffraction (RHEED) further developed simple PLD into laser-MBE \cite{Gross2000}. This allowed many groups to grow thin films of highest quality approaching semiconductor standards \cite{Gross2000,Klein1999,Klein2000,Gupta1990,Klein:2002a,Klein:2002b,Philipp:2003a,Philipp:2003b,Reisinger:2004a}. In close analogy to GaAs/AlAs heteroepitaxy, it is possible to grow complex heterostructures \cite{Wiedenhorst:1999a,Wiedenhorst:2000a,Lu:2000a,Philipp:2002a,Lu:2005a,Lu:2006a,Opel2011} composed of different oxides on suitable substrates in a layer-by-layer or block-by-block mode \cite{Gross2000,Reisinger2003a}. We note that due to the ionic bond character in oxides, the different atomic layers in general are not charge neutral and, hence, the energetically most favorable growth unit often is a molecular layer which is composed of one or several atomic layers to preserve charge neutrality. Therefore, molecular layer or block layer epitaxy is established for most oxides, whereas atomic layer epitaxy is common for semiconductor superlattice growth. This has been observed for the growth of MgO, Sr$_2$RuO$_4$, Fe$_3$O$_4$, and Sr$_2$CrWO$_6$ using stoichiometric targets \cite{Reisinger2003a,Reisinger2003b,Majewski2005,Geprags2013}. Due to the complex unit cells of MgO and Sr$_2$RuO$_4$, which consist of identical stoichiometric layers shifted by lattice translations, two RHEED oscillations per unit cell are observed. For Fe$_3$O$_4$, the unit cells consist of even four stoichiometric sub-units, resulting in four RHEED oscillations per unit cell.

In this Topical Review, we report on the recent developments of laser-MBE using the example of ZnO thin films and ZnO-based heterostructures. The Topical Review is organized as follows: In section~\ref{sec:technical}, we describe the technical realization of a laser-MBE setup with particular emphasis on the system developed at WMI. In section~\ref{sec:ZnO thin films}, the optimization of the deposition parameters for high-quality ZnO thin films is presented. In section~\ref{sec:heterostructures}, we extend the fabrication process to ZnO-based spintronic heterostructures which are important for spin transport experiments. Finally, in section~\ref{sec:summary} we summarize the key achievements.

\section{Technical Realization} \label{sec:technical}

In several previous publications \cite{Gross2000,Opel2012}, we have already described our laser-MBE system in detail. Therefore, we only give a brief overview on its technical realization, and then focus on a few important features in the subsections \ref{sec:optics} to \ref{sec:source}.

\begin{figure}[tb]
    \centering{\includegraphics[height=5.5cm]{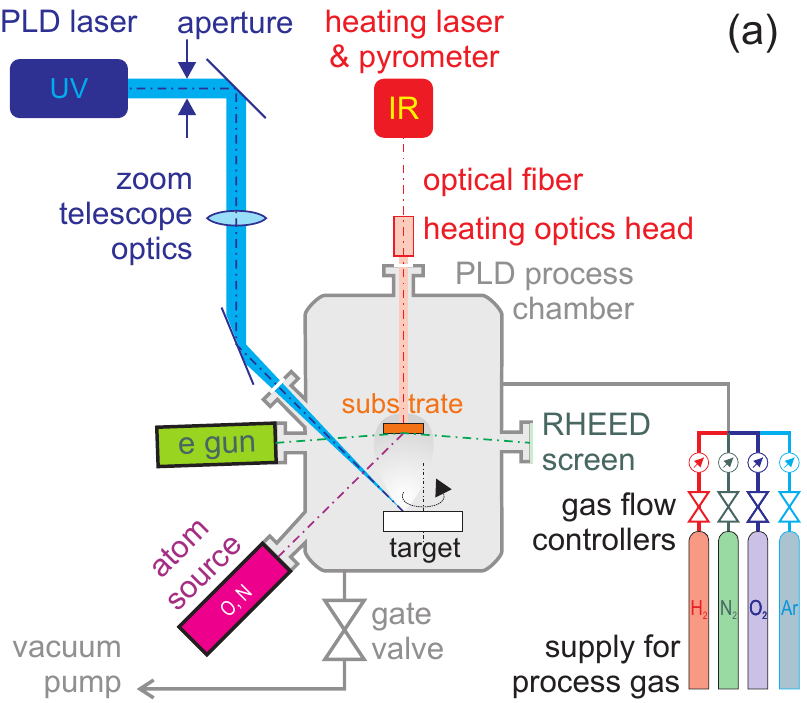}
    \hspace*{5mm}
    \includegraphics[height=5.5cm]{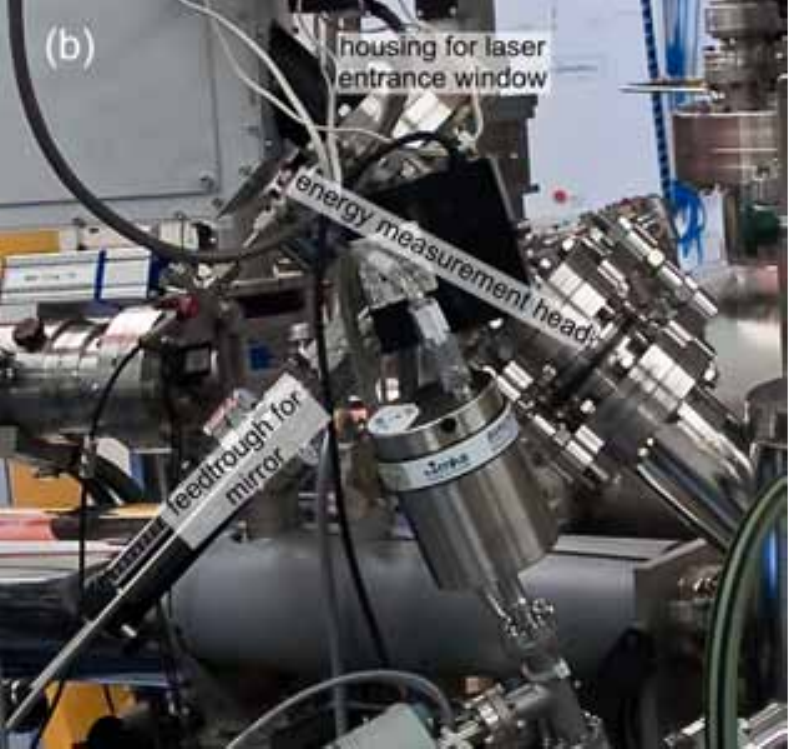}}
    \caption{\label{fig:technical}
             (a) Schematic of the UHV cluster system, operated at Walther-Mei{\ss}ner-Institut (WMI).
             (b) Photograph of the PLD chamber with UV laser entrance port assembly.}
\end{figure}

The schematic setup of our ultra-high vacuum (UHV) laser-MBE system is sketched in Figure~\ref{fig:technical}. The substrate is placed in a process chamber and heated from the back via infrared (IR) light from a continuous wave (cw) diode laser. The temperature is determined also from the back via a pyrometer which is placed outside the chamber. Both the IR heating laser and pyrometer are connected via two optical fibers to the heating optics head at the top entrance window of the process chamber. Further details are discussed below in subsection~\ref{sec:heater}. The process chamber is evacuated by an UHV pump and the process pressure is controlled via a gate valve using a down-stream pressure control. The composition of the process atmosphere is established by several gas flow controllers at the process gas inlet side. The chamber can be operated with reducing, inert, or oxidizing atmospheres (see subsection~\ref{sec:source}). In addition, an atom source is installed, providing additional radicals of oxygen or nitrogen (cf. subsection~\ref{sec:source}). The deposition process is monitored \textit{in situ} via a high-pressure reflection high-energy electron diffraction (RHEED) system. It consists of a differentially pumped electron gun and a fluorescent screen to visualize the electron interference pattern \cite{Klein1999,Klein2000}. The polycrystalline, stoichiometric target is placed in the lower part of the process chamber. It is rotated and irradiated off-center by the ultraviolet (UV) laser pulses generated by the external excimer laser. This leads to the formation of a stoichiometric plasma (the so-called ``plume'') which transfers material from the target to the substrate. The optical path of the pulsed UV laser beam consists of an aperture which is projected onto the target surface via a telescope zoom optics. In this way, the laser fluence at the target can be varied by adjusting the reproduction ratio while keeping the laser pulse energy constant. In the following, we describe the UV laser optics in more detail.

\subsection{Optics for ultraviolet laser} \label{sec:optics}

The laser fluence $\rho_{\mathrm L}$ is of crucial importance for the laser-MBE process since it determines the ablation of material from the target and the creation of the plasma plume. It is an important parameter for the correct stoichiometric transfer of material to the substrate and typically has to exceed a certain threshold ranging from $1-3$\,J/cm$^2$ for a 25 ns pulse. For optimization of the deposition process, $\rho_{\mathrm L}$  has to be reproducibly adjusted and finetuned to its optimum value. The laser fluence is a function of the UV laser pulse energy, the losses caused by optical components like mirrors, lenses, or windows and absorption in the process gas, and finally the area of the laser spot on the surface of the target. The losses are given by the optical path and are more or less constant at a given wavelength provided that the optical elements are kept free of surface contamination. The pulse energy of the used UV laser model can be adjusted over some range according to the technical specifications. However, a stable pulse energy is usually achieved only after the laser has emitted a significant number of pulses. Therefore, a fast tuning of $\rho_{\mathrm L}$ by changing the pulse energy (e.g. when growing multilayers from different target materials) is not possible. A more appropriate approach is to vary $\rho_{\mathrm L}$ by adjusting the area of the laser spot while keeping the pulse energy at a constant value.

\begin{figure*}[tb]
    \centering{\includegraphics[width=0.9\textwidth]{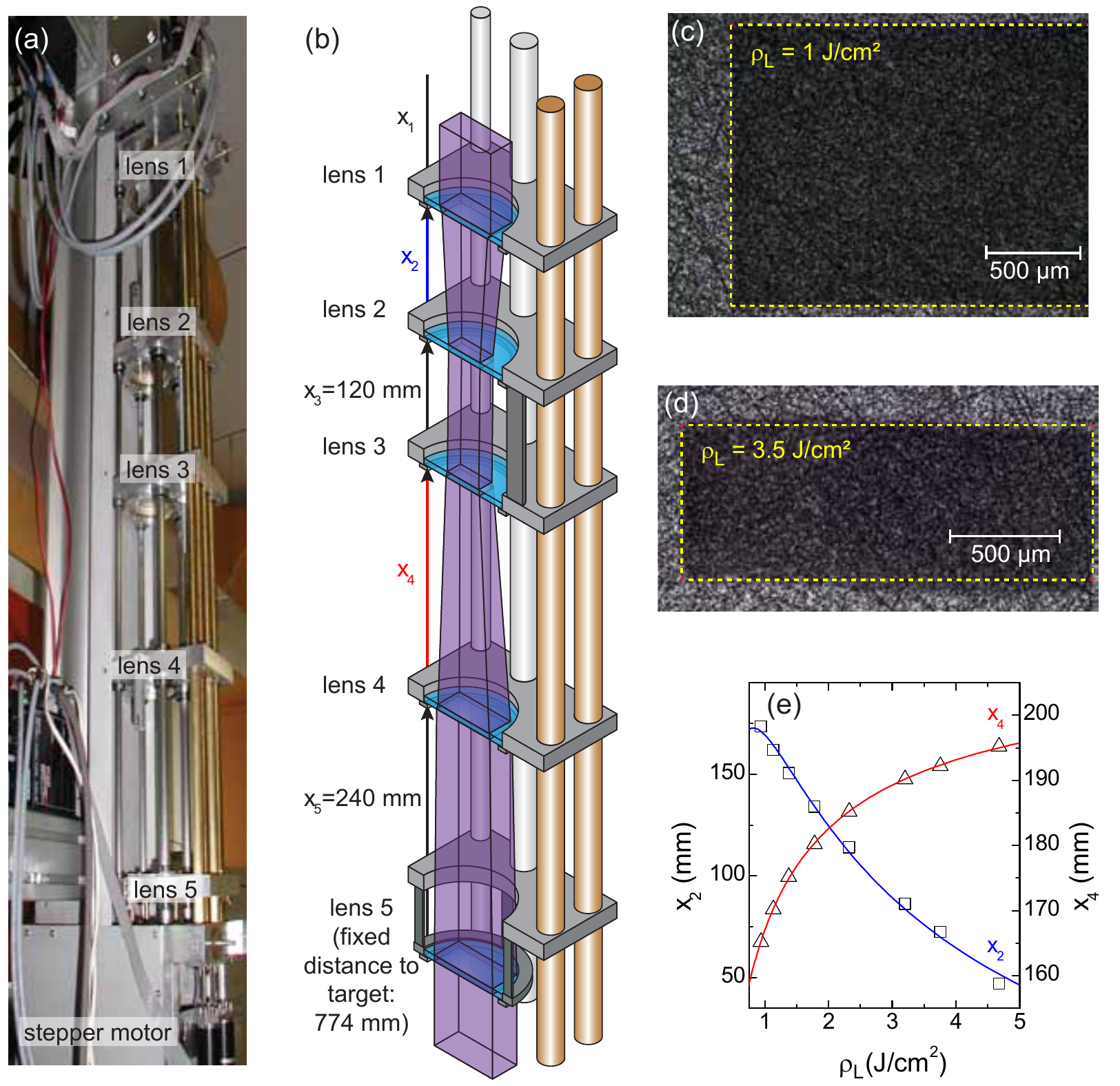}}
    \caption{\label{fig:optics}
             (a) Photograph and
             (b) Illustration of the telescope zoom optics consisting of five lenses. The lenses 1 to 5 are mounted on five lens holders.
             (c,d) Optical micrographs of the focal spot area (indicated by the dashed yellow rectangles) on a polycrystalline SrTiO$_3$ target obtained with the same laser pulse energy. The two different areas correspond to two different laser fluences $\rho_{\mathrm L}$ of (c) 1~J/cm$^2$ and (d) 3.5~J/cm$^2$.
             (e) Lens positions $x_2$ (blue, left scale) and $x_4$ (red, right scale) as a function of the desired laser fluence $\rho_{\mathrm L}$.}
\end{figure*}

For a controlled variation of the laser spot area, we developed a telescope zoom optics for the pulsed UV laser light (cf. Figure~\ref{fig:optics}). It consists of in total five lenses with different focal lengths of $350$\,mm, $-100$\,mm, $-100$\,mm, $350$\,mm, and $750$\,mm [Figure~\ref{fig:optics}(a)]. The lens holders are attached to independent stepper motors, each connected to a controller providing a positioning precision of less than 0.005\,mm. The controllers are driven via a PC, thus allowing for a full automation of the lens system itself. As sketched in Figure~\ref{fig:optics}(b), the lenses~1 to 5 are mounted on sliding lens holders allowing for a movement over a total distance of 1200\,mm.  The telescope zoom optics creates a sharp image of a rectangular aperture, which cuts out the central, spatially homogeneous part of the beam profile, on the surface of the target material [cf. Figure~\ref{fig:optics}(c,d)]. Note that the generation of a sharp and homogeneous beam profile is crucial to achieve a homogeneous laser fluence on the target. In turn, this results in a uniform ablation of material from the target and to a minimization of droplet formation \cite{Reisinger2003b}. In our zoom optics, the positions of lenses 1 to 3 determine the reduction ratio and hence the size of the image, while lenses 4 and 5 assure perfect focusing. The position of lens~5 is fixed at a distance of 774\,mm to the target in the process chamber. To reduce the number of parameters for the optimization of the focal point of the system for any given spot area, we keep the distances between lenses~2 and 3 fixed at $x_3 = 120$\,mm and that between lenses~4 and 5 at $x_5 = 240$\,mm. Only the distances $x_2$ or $x_4$ between lenses~1 and 2 or between lenses~3 and 4, respectively, are varied. Note that $x_1$ is simply determined from the other distances as the total length of the optical axis from the excimer laser to the target is given by 4020\,mm. With this telescope zoom optics, we are able to change the area of the UV laser spot on the target from 12\,mm$^2$ to 1.2\,mm$^2$, resulting in an accessible range of laser fluences from $\rho_{\mathrm L} = 0.5$\,J/cm$^2$ to 5\,J/cm$^2$. Figure~\ref{fig:optics}(e) displays the necessary lens positions $x_2$ and $x_4$ as a function of the desired laser fluence $\rho_{\mathrm L}$.

An omnipresent problem is the increase of losses along the optical path due to degradation of the surface coatings or contaminations of the different optical elements. Here, the main contribution usually results from absorption in the laser entrance port of the process chamber, since this window is continuously covered with material from the plasma plume during the PLD process. In order to maintain a stable laser fluence at the target, we have to compensate for this absorption effect by increasing the laser energy or by occasionally cleaning the window. We use a so-called \emph{intelligent} window (PVD Products) at the laser entrance port combining two unique features. First, it keeps the inner side of the entrance window free of coatings by blocking the ablated plasma plume via a rotatable disc consisting of UV grade fused silica. Once the exposed section of this disc is contaminated too much, it is simply rotated to expose an optically clean section gain. Second, an insertable mirror positioned in the light path after the disc allows to guide the incoming UV laser pulse through a side window, where its energy is determined by a pyroelectric detector with a resolution of $3\,\mu$J. These measures help to improve the deposition processes by accurately monitoring $\rho_{\mathrm L}$ as one of the most critical process parameters. Recording and adjusting the energy that actually arrives on the target before each run or even during the laser-MBE process results in a constant on-target laser fluence and helps to yield reproducible film properties and deposition rates (cf. subsection \ref{sec:fluence}).

\subsection{Substrate heating and temperature control} \label{sec:heater}

The realization of an \emph{epitaxial} thin film growth crucially depends on the correct deposition temperature, i.e.~the temperature $T_\mathrm{sub}$ of the substrate. For oxides, $T_\mathrm{sub}$ typically ranges from $300^\circ$C to more than $1000^\circ$C. Conventional resistive heaters located inside the UHV process chamber are problematic since the heating effect is not restricted to the substrate alone. Moreover, the use of resistive radiation heating unavoidably results in mechanical parts (such as the substrate carrier or the manipulator) with temperatures well above $T_\mathrm{sub}$. Their hot surfaces in turn limits the achievable base pressures. To overcome this problem, recently contact-less infrared (IR) optical heating systems have been developed. Such a system has first been introduced in 1999 by Ohashi \textit{et al.} and reached $T_\mathrm{sub} = 1450^\circ$C \cite{Ohashi1999}. The IR laser is positioned outside the vacuum system and the IR light is fed into the vacuum chamber via an optical fiber. A specific advantage of an IR laser heating is the fact that the laser can be switched on and off within a short time scales. This allows for rapid changes of $T_\mathrm{sub}$, since the substrate exhibits a significantly lower heat capacity than any resistive heater \emph{plus} substrate. In this way rapid temperature modulation epitaxy becomes possible \cite{Tsukazaki2005}. Furthermore, high temperatures up to the melting point of typical substrates like, e.g., Si ($1410^\circ$C) can be reached without using an extensive amount of radiation shielding preventing the optical access to the sample surface. IR laser heating, however, does not work with transparent substrates such as Al$_2$O$_3$. To overcome this problem, we deposit a thin metallic layer of Pt (typically $100-200$\,nm) on the back side of those substrates in order to increase the absorption of the IR laser light.

State-of-the art heating systems combine an IR laser with a pyrometer, both positioned outside the process chamber, for non-contact heating, temperature measurement, and temperature control of the substrate. The main part of our heating system (Surface LH-140) consists of a watercooled IR laser diode array with a maximum output power of 140\,W at a wavelength of 940\,nm. The IR laser diode array is mounted together with its power supply and a micro-PC with pyrometer card into the main control unit. This unit is connected to the process chamber via two optical fibers, one for the IR heating power and the second for the temperature measurement. A specially designed optics with a beam splitter attached to the process chamber focuses the IR laser beam onto the back side of the substrate and simultaneously collects its thermal radiation which is sent back to the control unit. There, a dual wavelength pyrometer measuring at 1711\,nm and 1941\,nm determines $T_\mathrm{sub}$ without the necessity to correct for the emittivity of the substrate. The temperature range of the pyrometer is $300^\circ$C to $1400^\circ$C in the one-colour mode and $500^\circ$C to $1400^\circ$C in the two-colour mode. The micro-PC allows for several control modes of the laser heating system, including programming, storing, and executing complex temperature vs.~time sequences.

The combination of IR laser heating and pyrometry makes the system very convenient as there are no parts inside the process chamber which need maintenance. We reach $T_\mathrm{sub} > 1200^\circ$C at a vacuum in the $10^{-8}$\,mbar range. Compared to the previously installed radiation heater, an improvement of the background pressure of about one order of magnitude is achieved \cite{Reisinger2003b}.  We also realized temperature sweeps of up to 500\,K/min, only limited by the mechanical stability of the respective substrate. In summary, the flexible heating system in combination with different gas atmospheres offers a wide range of process conditions for the thin film deposition. As an example, we discuss the dependence of the structural quality of ZnO thin films on the deposition temperature in subsection~\ref{sec:growth-temperature}.

\subsection{Target materials and process atmosphere} \label{sec:source}

In most cases polycrystalline targets are used in the PLD process. However, also single crystals can be used. The materials in the target can be supplemented by atomic (e.g. Ar) and molecular gases (e.g. O$_2$, N$_2$) forming the background process atmosphere or by atomic species (e.g. O, N) from independent radical sources operated during deposition. The target material usually consists of a pressed and sintered polycrystalline pellet with the desired stoichiometric composition. It must show a high density and little porosity to avoid the ejection of macroparticles when being hit by the UV laser pulse. A uniform ablation from the target can be ensured by scanning the laser beam across its surface while simultaneously rotating the target. For reproducible growth conditions, effects of target contamination e.g.~via re-deposition from the plasma have to be considered and if possible avoided.

Our process chamber contains a horizontal target carousel, capable of five different targets with maximum diameters of 1~inch. Two stepper motors allow for the PC-controlled rotation of the targets and the carousel, respectively. The whole carousel can be shifted in vertical direction via a third stepper motor to correct for different thicknesses of the respective targets. The correct vertical target position is determined by the intercept of the vertical shift axis with the optical axis of the zoom optics discussed above in subsection \ref{sec:optics}. Furthermore, the correct substrate position is determined by the focus of the RHEED electron beam. Therefore, in our system the target-to-substrate distance is fixed at 60\,mm. For reproducible deposition conditions we carry out a pre-ablation prior to each thin film deposition process and mechanically clean the targets in regular time intervals.

As discussed above, pressure and composition of the deposition atmosphere (inert, oxidizing, reducing) are important process parameters. For the fabrication of complex oxide heterostructures by subsequent deposition of different materials from different targets, both the pressure and the composition of the respective process gases have to be changed independently and quickly (cf. subsection~\ref{sec:heterostructures}). Our process gas supply contains N$_2$, Ar (both inert), O$_2$ (oxidizing), and 5\% H$_2$ in Ar (reducing), all of them connected to the process chamber via four independent gas flow controllers. The base pressure of our UHV process chamber is in the 10$^{-9}$\,mbar regime. The composition of the process atmosphere is set via mass flow controllers. The UHV pump is connected to the process chamber via a  motor-driven gate valve enabling a down-stream pressure control from about 1\,mbar down to $10^{-4}$\,mbar with a resolution of down to $10^{-5}$\,mbar. For even higher pressures, a bypass pumping line with smaller adjustable cross-sectional area is available. It allows for stable pressures in the range of $10^{-3}$\,mbar to 10\,mbar with a resolution of $10^{-3}$\,mbar. Combing different atmospheres at different pressures and temperatures can also be used for the annealing of substrates and thin films before, during, or after deposition. Examples for the effect of annealing are discussed in subsection~\ref{sec:process-buffer}.

In some cases, e.g.~for nitrogen doping of ZnO thin films, the use of atomic instead of molecular species is required. Therefore, we attached a radio frequency (rf) atom radical source to the laser-MBE chamber. This source can be used with N$_2$, O$_2$, and H$_2$ gases. It is water-cooled and UHV compatible. It operates with a maximum power of 600\,W at an rf frequency of 13.56\,MHz. It is equipped with an optical emission detector together with a plasma controller and an automatic tuning unit to allow for stable operation on a timescale of several hours. Its influence on the ZnO thin film growth is discussed below in subsection~\ref{sec:atom-source}.

\section{Deposition of ZnO Thin Films} \label{sec:ZnO thin films}

In this section we present the application of our laser-MBE system to the growth of epitaxial ZnO thin films and ZnO based heterostructures. The fabrication of highly oriented ZnO thin films by laser deposition was first reported in 1983 \cite{Sankur1983}. Here, we describe our more advanced deposition process and discuss the influence of the various process parameters on the structural properties of the resulting ZnO thin films.

\subsection{Substrate issues} \label{sec:substrates}

Bulk ZnO crystallizes in the polar, hexagonal wurtzite structure (point group $C_{6v}$ or $6mm$, space group $C^4_{6v}$ or $P6_3mc$) with lattice parameters $a = 0.32501$\,nm and $c = 0.52071$\,nm (see Table~\ref{tab:substrates}) \cite{Kisi1989}. ZnO thin films have already been deposited on many different substrate materials including Si \cite{Sankur1983,Lad1980}, GaAs \cite{Sankur1983}, Al$_2$O$_3$ \cite{Sankur1983,Chen1998,Kaidashev2003,Chen2005,El-Shaer2005,Nielsen2006}, ZnO \cite{Tsukazaki2010,Chen2005,Nielsen2006,Heinze2007,Neumann2007}, and ScAlMgO$_4$ \cite{Tsukazaki2005,Wessler2002} with varying results regarding their structural, electrical and optical quality. The relevant in-plane lattice mismatch for $c$-axis oriented growth on hexagonal substrates is summarized in Table~\ref{tab:substrates}. For the studies in this work, we focus on the heteroepitaxial growth on (0001)-oriented Al$_2$O$_3$ ($c$-plane sapphire) substrates. In contrast to the homoepitaxial growth on ZnO, this heteroepitaxial approach enables us to clearly separate film and substrate physical properties, however, at the expense of the formation of dislocations in the ZnO film due to the large lattice mismatch.

\begin{table}[tb]
    \caption{\label{tab:substrates}In-plane lattice parameter $a$ and in-plane lattice mismatch $\Delta a/a$ for ZnO on different substrates.}
    \begin{indented}
    \lineup
    \item[]\begin{tabular}{@{}*{7}{l}}
    \br
    &ZnO&Al$_2$O$_3$&ScAlMgO$_4$\cr
    \mr
    $a$\,(nm) & $0.32501$ & \0$0.4759$ & $0.3246$\cr
    $\Delta a/a$ & -- & $18.2{\%}^{\rm a}$ & $0.126{\%}$\cr
    \br
    \end{tabular}
    \item[] $^{\rm a}$ the ZnO lattice is rotated by $30^\circ$
    \end{indented}
\end{table}

\begin{figure}[b]
    \centering{\includegraphics[width=5cm]{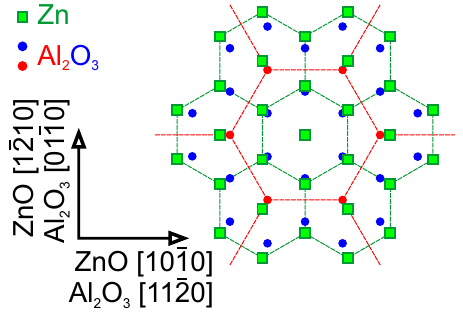}}
    \caption{\label{fig:ZnO-Al2O3}
             In-plane epitaxial relationship between the hexagonal lattices of Al$_2$O$_3$ and ZnO, rotated by $30^\circ$ (according to  \cite{Ozgur2005}).}
\end{figure}

The $c$-plane of Al$_2$O$_3$ consists of alternating layers of O (with 6-fold symmetry) and Al (3-fold symmetry). In contrast, the wurtzite structure of ZnO shows a 6-fold symmetry for both Zn and O in the (0001) planes. When growing ZnO on top of Al$_2$O$_3$ ($a=0.4759$\,nm), the close-packed O sublattices of ZnO and Al$_2$O$_3$ align themselves \cite{Chen1998}, in analogy to (111)-oriented Fe$_3$O$_4$ thin films on (0001)-oriented ZnO or Al$_2$O$_3$ substrates \cite{Boger2008}. This results in a rotation of the ZnO unit cell by $30^\circ$ around the (0001) direction (cf. Figure~\ref{fig:ZnO-Al2O3}). The epitaxial relationship between substrate and thin film therefore is $\mathrm{ZnO}[10\overline{1}0]\|\mathrm{Al}_2\mathrm{O}_3[11\overline{2}0]$ and $\mathrm{ZnO}[1\overline{2}10]\|\mathrm{Al}_2\mathrm{O}_3[01\overline{1}0]$, as shown in Figure~\ref{fig:ZnO-Al2O3}. Since the in-plane O-O distance in Al$_2$O$_3$ is given by $a^\prime =0.275$\,nm, this effectively reduces the in-plane lattice mismatch from $-30.4\%$ to $18.2\%$. However, this still large misfit and the resulting high dislocation density dominates the physical properties of the ZnO thin film. We note that due to the bulk inversion asymmetry of the wurtzite structure (i.e.~the $[0001]$ and $[000\overline{1}]$ directions are not equivalent) the alignment of the Zn sublattice of ZnO to the O sublattice of Al$_2$O$_3$ leads to the orientation of the [0001] direction pointing towards the substrate, implying that the electric polarization in ZnO points toward the film surface \cite{Tampo2008}.

\subsection{Optimization of the growth parameters}

The results presented in the following were obtained from epitaxial ZnO thin films deposited by laser-MBE, using a pulsed KrF excimer laser (248\,nm) with a pulse repetition rate of 2\,Hz. All thin films were grown in a pure O$_2$ atmosphere from the same  stoichiometric ZnO target, consisting of a pressed and sintered pellet of commercially available ZnO powder. The structural quality was then characterized by high-resolution X-ray diffractometry (HR-XRD) and reflectometry (HR-XRR) in a four-circle diffractometer (Bruker D8 Discover) using Cu $K\alpha_1$ radiation.

\subsubsection{Laser fluence.} \label{sec:fluence}

\begin{figure*}[tb]
   \centering{ \includegraphics[width=0.95\textwidth]{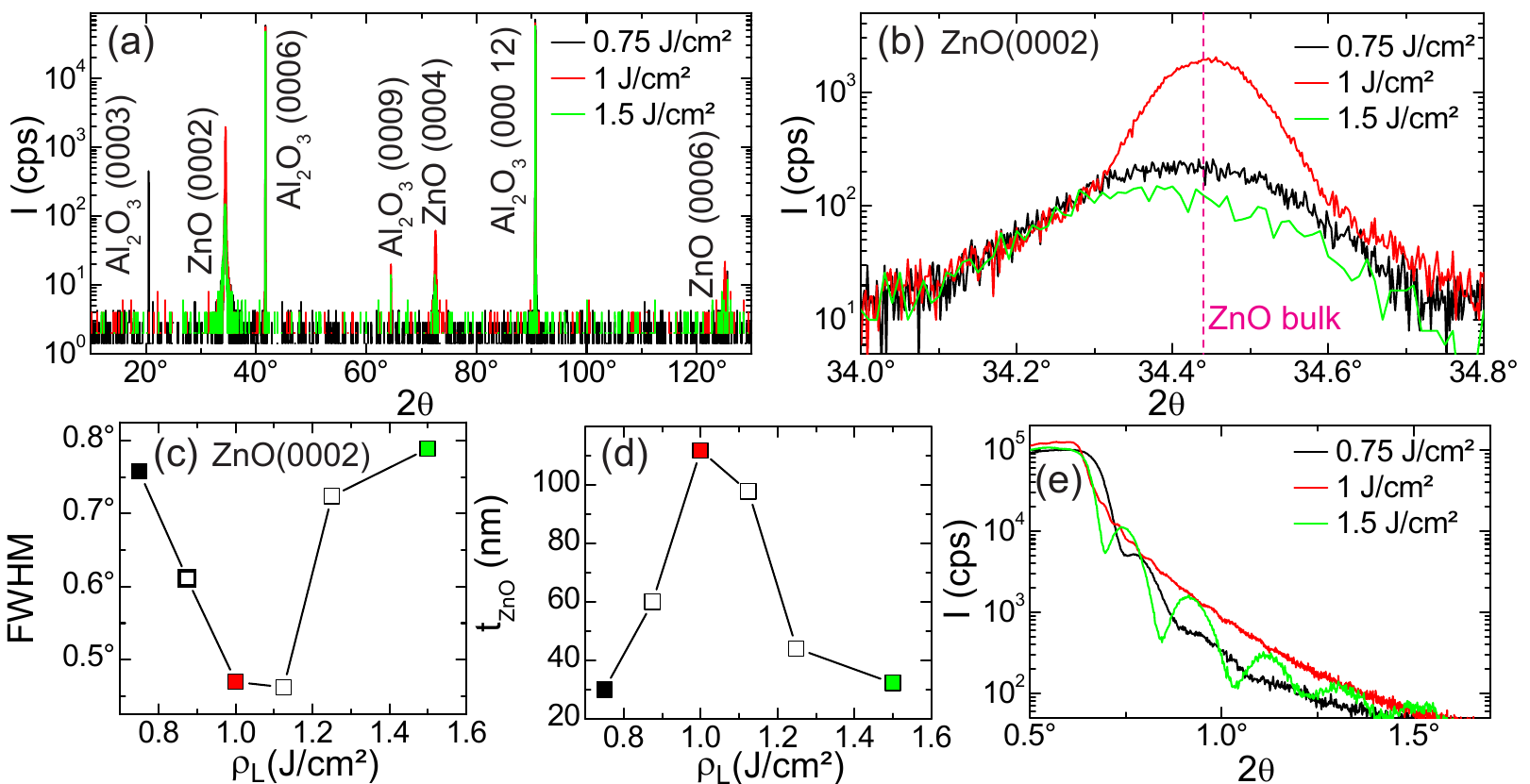}}
    \caption{\label{fig:fluence}
             HR-XRD results for ZnO films deposited on (0001)-oriented Al$_2$O$_3$ substrates at different values of the laser fluence $\rho_\mathrm{L}=0.75$~J/cm$^2$ (black), 1~J/cm$^2$ (red), and 1.5~J/cm$^2$ (green). For all films we used 2000 laser pulses,  $T_\mathrm{sub} = 600^\circ$C, and  $p_{\rm O_2} = 10^{-3}$\,mbar.
             (a) Out-of-plane $\omega$-$2\theta$ diffraction scans.
             (b) Out-of-plane $\omega$-$2\theta$ diffraction scans on an enlarged scale around the ZnO(0002) reflection. The position of the reflection for bulk ZnO is indicated by the vertical dashed line.
             (c) Full widths at half maximum (FWHM) of the $\omega$ rocking curves of the ZnO(0002) reflection.
             (d) Thicknesses $t_\mathrm{ZnO}$ of the ZnO thin films.
             (e) X-ray reflectometry scans.}
\end{figure*}

We first determine the optimum laser fluence $\rho_\mathrm{L}$, for which an optimum in structural properties is obtained. We therefore deposited a set of six ZnO thin films using 2000 laser pulses, a constant substrate temperature of $T_\mathrm{sub} = 600^\circ$C and a constant oxygen pressure of  $p_{\rm O_2} = 10^{-3}$\,mbar, while varying the laser fluence $\rho_\mathrm{L}$ from 0.75 to 1.5\,J/cm$^2$. The experimental results of the structural characterization via HR-XRD and HR-XRR are summarized in Figure~\ref{fig:fluence}.

As shown by Figure~\ref{fig:fluence}(a), the out-of-plane $\omega$-$2\theta$ scans display only reflections from either the ZnO thin film or the Al$_2$O$_3$ substrate and do not reveal any secondary phases. The positions of the ZnO(0002) reflections for all films are close to the ZnO bulk value of $2\theta = 34.44^\circ$ [cf. Figure~\ref{fig:fluence}(b)]. This suggests a relaxed film growth due to the large lattice mismatch and has been observed for all films on Al$_2$O$_3$. The ZnO(0002) reflection shows maximum intensity for the films grown with $\rho_\mathrm{L} = 1$\,J/cm$^2$ or 1.25\,J/cm$^2$. Figure~\ref{fig:fluence}(c) shows the full width at half maximum (FWHM) of the $\omega$ rocking curves around the ZnO(0002) reflection. It has a minimum of $\sim 0.45^\circ$ for $\rho_\mathrm{L} = 1$\,J/cm$^2$ or 1.25\,J/cm$^2$ indicating a small out-of-plane mosaic spread, but increases towards $0.8^\circ$ for lower or higher $\rho_\mathrm{L}$. In spite of the same number of laser pulses, the thickness $t_\mathrm{ZnO}$ of the six films differs by almost a factor of~4 [see Figure~\ref{fig:fluence}(d)]. The largest $t_\mathrm{ZnO} = 110$\,nm is observed for $\rho_\mathrm{L} = 1$\,J/cm$^2$. The values for $t_\mathrm{ZnO}$ were extracted from fits to the corresponding HR-XRR spectra shown in Figure~\ref{fig:fluence}(e). From HR-XRR, we also see that the about exponential intensity decay with increasing angle, which is indicative for the film roughness, is lowest for $\rho_\mathrm{L} = 1$\,J/cm$^2$. In addition, for this fluence the oscillation period of the beating pattern superimposed on the exponential decay is smallest due to the larger film thickness. For decreasing $\rho_\mathrm{L}$, less target material is ablated per pulse. Therefore, at the same pulse number the film thickness is expected to decrease with decreasing $\rho_\mathrm{L}$, in agreement with the data for lower fluences. For high $\rho_\mathrm{L}$, not only the amount of target material ablated per pulse but also the kinetic energy of the plasma material is increasing. The latter is likely to cause re-sputtering effects and a reduction of the sticking coefficients. As shown by Figure~\ref{fig:fluence}(d), this results in a decrease of the film thickness when we are increasing $\rho_\mathrm{L}$ beyond 1\,J/cm$^2$. We also note that with increasing laser fluence and hence increasing plume density the self-absorption of the laser is increasing. In summary, the optimization of the laser fluence yields an optimum value of $\rho_\mathrm{L} = 1$\,J/cm$^2$ for the growth of ZnO thin films with high growth rate and satisfactory structural properties.

As pointed out above, the epitaxial in-plane relationship between ZnO and Al$_2$O$_3$ depends on the relative alignment of ZnO to the O-sublattice of Al$_2$O$_3$. In recent publications \cite{Yoshimoto1995,Akiyama2007}, it has been demonstrated that a substrate annealing step at elevated temperatures prior to deposition leads to an oxygen-terminated surface, improving both the quality of the substrate and the epitaxial ZnO thin film. We studied this effect and found a significant improvement of the structural quality after annealing the substrate at $T_\mathrm{sub} = 850^\circ$C for 1\,h in a pure O$_2$ atmosphere of $10^{-3}$\,mbar. For the data shown below, all Al$_2$O$_3$ substrates have been annealed \textit{in-situ} prior to ZnO thin film deposition.

\subsubsection{Substrate temperature.} \label{sec:growth-temperature}

\begin{figure*}[tb]
    \centering{\includegraphics[width=0.95\textwidth]{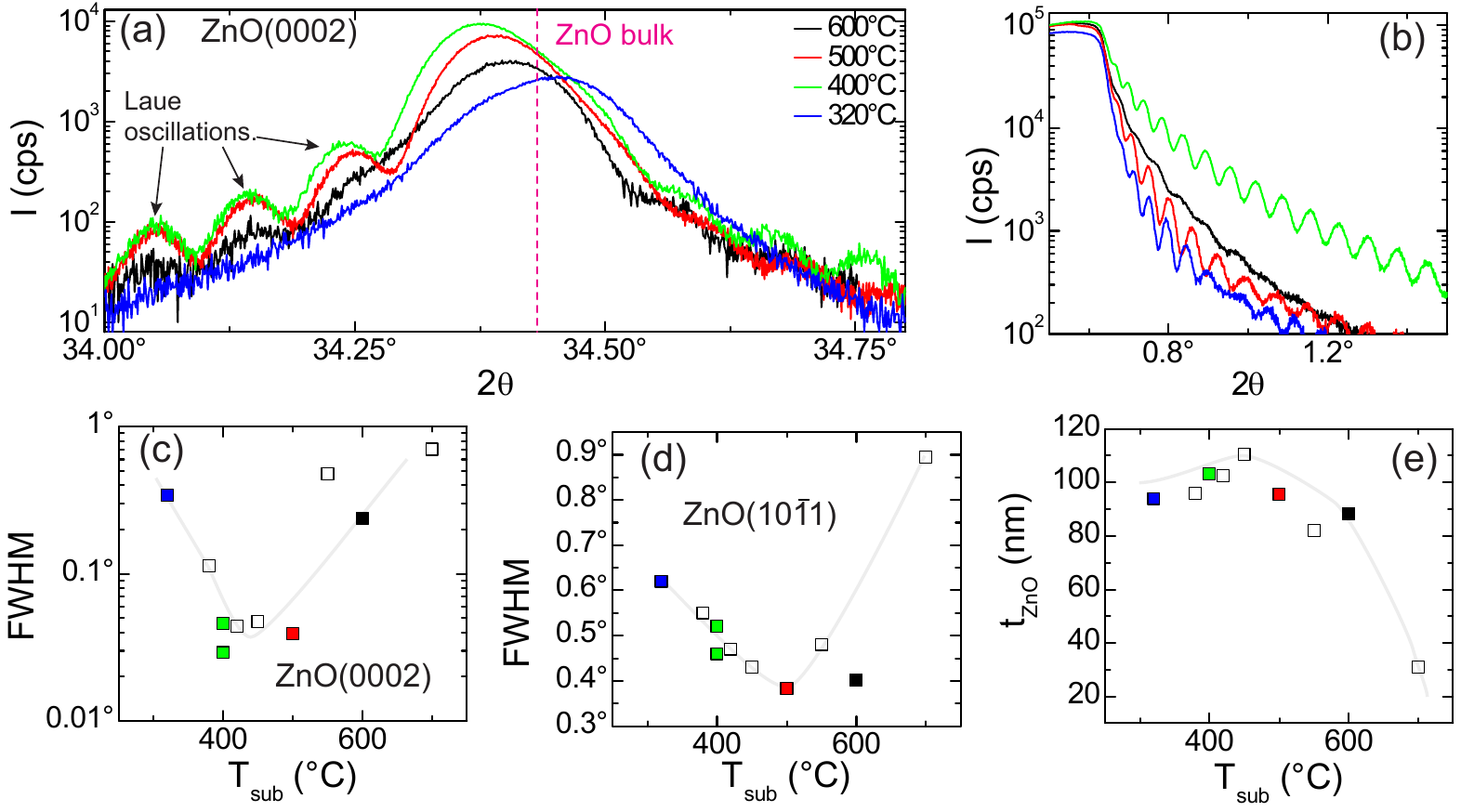}}
    \caption{\label{fig:temperature}
             HR-XRD results for ZnO films deposited on (0001)-oriented Al$_2$O$_3$ substrates at different values of the substrate temperature $T_\mathrm{sub}=320^\circ$C (blue), $400^\circ$C (green), $500^\circ$C (red), and $600^\circ$C (black). For all films we used 2000 laser pulses, $\rho_\mathrm{L}=1$~J/cm$^2$, and  $p_{\rm O_2} = 10^{-3}$\,mbar.
             (a) Out-of-plane $\omega$-$2\theta$ diffraction scans on an enlarged scale around the ZnO(0002) reflection. The position of the reflection for bulk ZnO is indicated by the vertical dashed line.
             (b) X-ray reflectometry scans.
             (c) Full widths at half maximum (FWHM) of the $\omega$ rocking curves of the ZnO$(0002)$ reflection.
             (d) FWHM of the $\omega$ rocking curves of the ZnO$(10\overline{1}1)$ reflection.
             (e) Film thickness $t_\mathrm{ZnO}$ of the ZnO thin films deposited at different $T_\mathrm{sub}$.
             The light grey lines in (c,d,e) are guides to the eye.}
\end{figure*}

We next determine the optimum substrate temperature $T_\mathrm{sub}$. We fabricated a second set of ZnO thin films on \textit{in-situ} annealed Al$_2$O$_3$ substrates at various substrate temperatures $T_\mathrm{sub}$ ranging between $320^\circ$C and $700^\circ$C, while using constant values for the laser fluence ($\rho_\mathrm{L} = 1$\,J/cm$^2$), the number of laser pulses (2000), and the oxygen pressure ($p_{\rm O_2} = 10^{-3}$\,mbar). The experimental results of the structural characterization via HR-XRD and HR-XRR are summarized in Figure~\ref{fig:temperature}.

Again, the out-of-plane $\omega$-$2\theta$ scans display only reflections from either the thin film or the substrate and do not reveal any secondary phases. Figure~\ref{fig:temperature}(a) shows the $\omega$-$2\theta$ scans around the ZnO(0002) reflection. Evidently, we now see satellites due to Laue oscillations for $400^\circ\mathrm{C} \leq T_\mathrm{sub} \leq 500^\circ\mathrm{C}$, reflecting coherent growth in the out-of-plane direction. In addition, the FWHM of the $\omega$ rocking curve for the out-of-plane ZnO(0002) reflection is significantly reduced compared to films grown at $600^\circ$C shown in Figure~\ref{fig:fluence}. As shown in Figure~\ref{fig:temperature}(c), the FWHM reaches a minimum value of $0.04^\circ$ for $400^\circ\mathrm{C} \leq T_\mathrm{sub} \leq 500^\circ\mathrm{C}$, indicating a low spread of the out-of-plane mosaic misorientation. The FWHM of the $\omega$ rocking curve for the asymmetric ZnO$(10\overline{1}1)$ reflection, indicative of both the in-plane (which strongly influences the mobility of charge carriers in ZnO on Al$_2$O$_3$) \cite{Miyamoto2002} and the out-of-plane mosaic misorientation, has its lowest value of $0.4^\circ$ at $T_\mathrm{sub} = 500^\circ$C [cf. Figure~\ref{fig:temperature}(d)]. To further narrow down the optimum growth temperature we analyze the HR-XRR for the various films. As shown in Figure~\ref{fig:temperature}(b), all curves exhibit pronounced oscillations which allow for the determination of the film thickness $t_\mathrm{ZnO}$ plotted in Figure~\ref{fig:temperature}(e). Obviously, the exponential decrease of the intensity with increasing angle is smallest for $T_\mathrm{sub} = 400^\circ$C. Finally, Figure~\ref{fig:temperature}(e) shows that the film thickness $t_\mathrm{ZnO} \simeq 100$\,nm is nearly independent of the deposition temperature except for the highest temperature $T_\mathrm{sub} = 700^\circ$C.

From this $T_\mathrm{sub}$-series we choose $T_\mathrm{sub} = 400^\circ$C as the optimum deposition temperature. In this context we note that a low substrate temperature is favored since it allows for the growth of sharp interfaces and reduces interdiffusion of Al from the Al$_2$O$_3$ substrate into the ZnO thin film. This is important to reduce the residual carrier concentration in ZnO because Al is known to form a shallow donor \cite{McCluskey2007}.

\subsection{Buffer layer system} \label{sec:process-buffer}

A further increase in sample quality can be achieved by introducing a buffer layer between the substrate and the ZnO thin film to reduce the lattice mismatch. Several groups reported on an improvement of structural and electrical quality of ZnO thin films with different complex buffer layer systems, including an annealed ZnO buffer \cite{Kaidashev2003,Koyama2007}, a (Mg,Zn)O buffer with different Mg content \cite{El-Shaer2005,Wassner2009}, and a double buffer layer of ZnO/MgO \cite{Miyamoto2002}. However, the introduction of a buffer layer also prevents the unambiguous determination of the physical properties of the ZnO thin film, since the contributions of film and buffer layer are difficult to separate. In order to reduce the complexity, we restrict our investigation to a high temperature-annealed ZnO buffer layer system.

\subsubsection{Optimum process script.}

\begin{figure*}[tb]
    \centering{\includegraphics[width=0.95\textwidth]{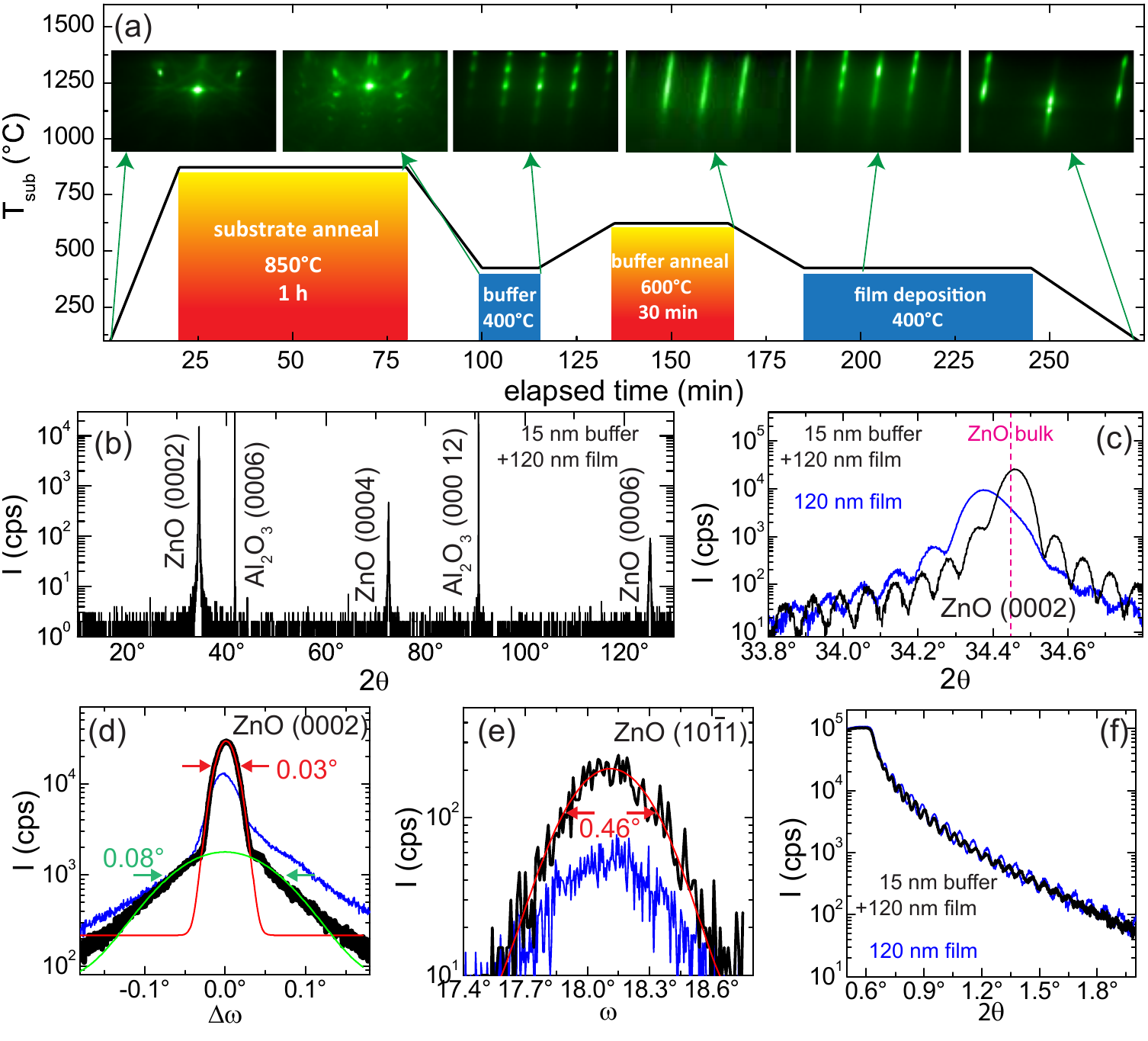}}
    \caption{\label{fig:process-buffer}
             Deposition of ZnO thin films with and without ZnO buffer layer on (0001)-oriented Al$_2$O$_3$ substrates.
             (a) $T_\mathrm{sub}$-profile for the deposition of a 120\,nm thin ZnO film with a 15\,nm thin buffer layer (black solid line). The annealing and growth intervals are shaded in red and blue, respectively. The insets show representative RHEED patterns at different stages before, during, and after the deposition process, indicated by green arrows.
             (b) Out-of-plane $\omega$-$2\theta$ diffraction scans.
             (c) Out-of-plane $\omega$-$2\theta$ diffraction scans on an enlarged scale around the ZnO(0002) reflection. The position of the reflection for bulk ZnO is indicated by the vertical dashed line.
             (d) Out-of-plane $\omega$ rocking curves around the ZnO(0002) reflection. The red and green lines represent two Gaussian fits to the data for the film with buffer layer, the FWHM is indicated by the arrows.
             (e) $\omega$ rocking curves around the ZnO$(10\overline{1}1)$ reflection. The red line represents a Gaussian fit to the data for the film with buffer layer.
             (f) X-ray reflectometry scans.}
\end{figure*}

We first deposit the ZnO buffer layer at the same conditions as the ZnO thin film ($\rho_\mathrm{L} = 1$\,J/cm$^2$, 2\,Hz, $T_\mathrm{sub} = 400^\circ$C, $p_{\rm O_2} = 10^{-3}$\,mbar). The whole process together with the data from HR-XRD and HR-XRR is summarized in Figure~\ref{fig:process-buffer}. The chronological evolution of the deposition including substrate annealing is depicted in Figure~\ref{fig:process-buffer}(a) together with the corresponding RHEED patterns. In a first step (1), the Al$_2$O$_3$ substrate is heated up to $T_\mathrm{sub} = 850^\circ$C at a rate of $30^\circ$C/min in $p_{\rm O_2} = 10^{-3}$\,mbar and then annealed for 60\,min. During the substrate annealing, additional diffraction peaks appear in the RHEED pattern and the sharpness of the pattern increases, indicating a surface reconstruction with oxygen termination. In the second step (2), $T_\mathrm{sub}$ is decreased to $400^\circ$C at $-30^\circ$C/min, and the ZnO buffer layer is grown with a typical thickness of 15\,nm. The RHEED pattern becomes spotty indicating a rough surface, and the spacings between the streaks increase. In the third step (3), the sample is heated up again to $T_\mathrm{sub} = 600^\circ$C at a rate of $10^\circ$C/min and annealed for 30\,min. This buffer annealing changes the RHEED pattern into a streaky one arising from the increased smoothness of the surface. In the fourth step (4), the temperature is lowered again to $T_\mathrm{sub} = 400^\circ$C at a rate of $10^\circ$C/min, and a 120\,nm thick ZnO film is deposited. After a few pulses, the streaks in the RHEED pattern are overlayed by spots, representing a three-dimensional growth of the ZnO thin film on the buffer layer. Finally, in the fifth step (5), the sample is cooled down to room temperature at a rate of $10^\circ$C/min in $p_{\rm O_2} = 10^{-3}$\,mbar. The RHEED pattern remains streaky with superimposed spots, indicating an increase of the surface roughness. For samples without buffer layer, the steps (2) and (3) are skipped.

\subsubsection{Structural characterization.} \label{sec:rsm}

The out-of-plane $\omega$-$2\theta$ scan from the buffered ZnO thin film in Figure~\ref{fig:process-buffer}(b) exhibits only reflections which can be either attributed to the substrate or to the film, indicating the absence of any secondary phases. The situation around the ZnO(0002) reflection is displayed in Figure~\ref{fig:process-buffer}(c) on an enlarged scale for a buffered (black) and an unbuffered thin film (blue) of the same thickness. Satellites due to Laue oscillations are visible for both samples. However, for the unbuffered sample, the (0002) reflection is slightly asymmetric and shifted away from the value for ZnO bulk, indicating an out-of-plane strain gradient in the sample. In contrast, the buffered ZnO thin film shows a symmetric (0002) reflection close to the bulk value. This result demonstrates that the introduction of an annealed buffer layer significantly reduces the strain in the ZnO film on top. As shown in Figure~\ref{fig:process-buffer}(d), the out-of-plane $\omega$ rocking curve of the ZnO(0002) reflection is asymmetric for the unbuffered sample and symmetric with higher intensity for the buffered one. Both curves show two components: a broad background peak is superimposed on a narrow one with higher intensity. A Gaussian fit to the data from the buffered sample yields $\mathrm{FWHM} = 0.08^\circ$ for the broad peak and $\mathrm{FWHM} = 0.03^\circ$ for the narrow one. We attribute the broad contribution to the ZnO buffer layer which has a slightly higher mosaic spread than the narrow ZnO top layer. The latter value is excellent and indicates a small out-of-plane mosaic spread of the ZnO film. In contrast, as shown by Figure~\ref{fig:process-buffer}(e) the $\omega$ rocking curve of the asymmetric ZnO$(10\overline{1}1)$ reflection indicates a larger in-plane mosaic spread of the film with buffer. From a Gaussian fit, we extract $\mathrm{FWHM} = 0.46^\circ$ for the buffered sample. Compared to results from other groups, however, both FWHM values for the symmetric and asymmetric reflections are on par or do even exceed previously published data for the growth of ZnO thin films on (0001)-oriented Al$_2$O$_3$ substrates \cite{Kaidashev2003,Miyamoto2002,Wassner2009}. Finally, from the XRR spectra shown in Figure~\ref{fig:process-buffer}(f) we extract a total film thickness of 135\,nm and a surface roughness below $0.5$\,nm for the buffered ZnO sample. These results demonstrate that the introduction of an annealed ZnO buffer layer results in a significant improvement of the structural quality of the subsequent ZnO film with low interface and surface roughness.

For ZnO thin films grown on (0001)-oriented Al$_2$O$_3$ by different deposition techniques, the existence of rotational domains has been reported \cite{Chen1998,Fons1999}. They have a significant influence on structural, electrical, and optical properties \cite{Chen1998}. From $\phi$-scans of the $\mathrm{Al}_2\mathrm{O}_3\{11\overline{2}6\}$ and $\mathrm{ZnO}\{10\overline{1}1\}$ diffraction peaks in samples with and without ZnO buffer layer, we find a 6-fold symmetry for both reflections (not shown here). We do not detect additional peaks for ZnO, indicating that no rotational domains are present in our thin film samples. Furthermore, the reflections from the substrate and the film coincide at the same $\phi$ values confirming the in-plane parallel alignment of the O-sublattices of ZnO and Al$_2$O$_3$, $\mathrm{ZnO}[10\overline{1}0]\|\mathrm{Al}_2\mathrm{O}_3[11\overline{2}0]$, as expected above (see Figure~\ref{fig:ZnO-Al2O3}).

\begin{figure*}[tb]
    \centering{\includegraphics[width=0.95\textwidth]{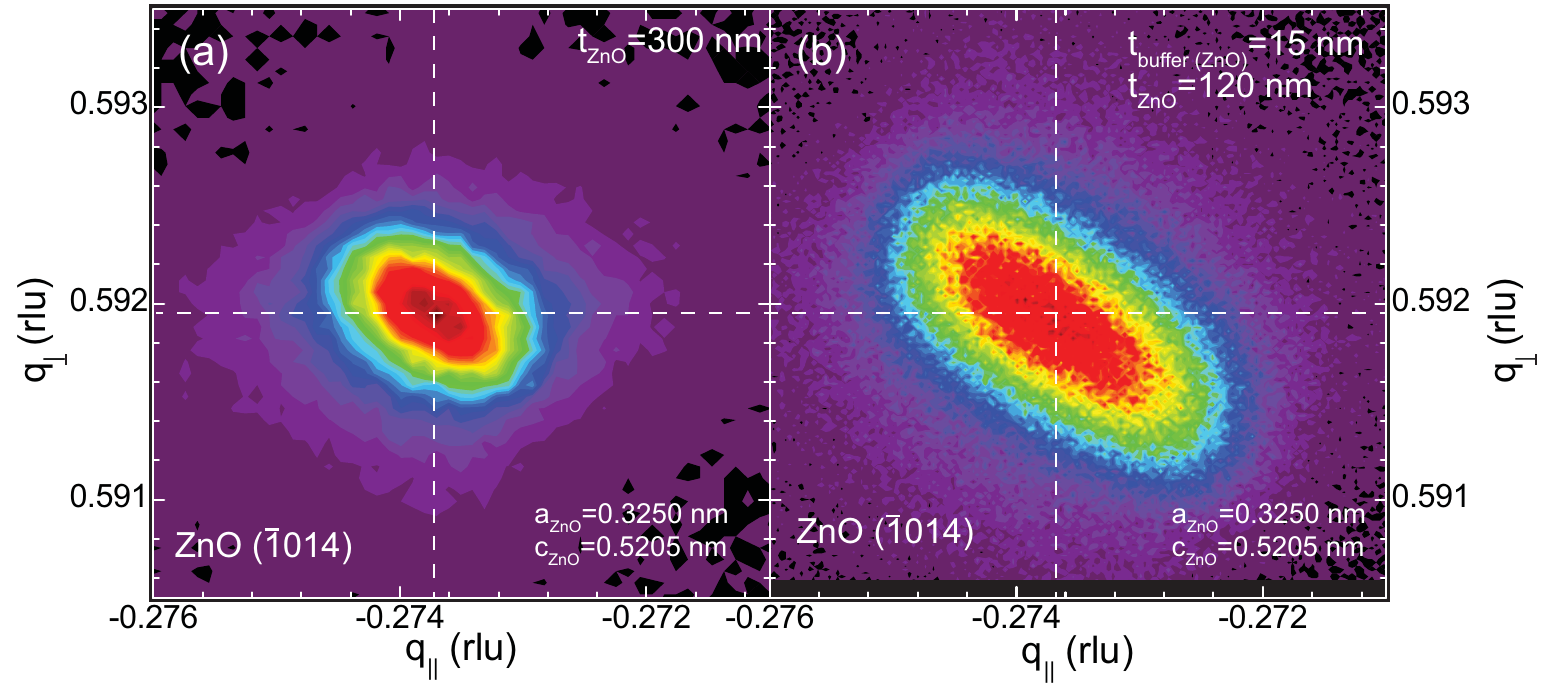}}
    \caption{\label{fig:rsm}
             X-ray reciprocal space maps (RSM) around the ZnO $(\overline{1}014)$ reflection.
             (a) ZnO thin film with a thickness of $t_\mathrm{ZnO}=300$\,nm on a (0001)-oriented Al$_2$O$_3$ substrate.
             (b) ZnO thin film ($t_\mathrm{ZnO}=120$\,nm) on a ZnO buffer layer ($t_\mathrm{buffer}=15$\,nm) on a (0001)-oriented Al$_2$O$_3$ substrate. The position of the maximum intensity is marked by dashed white lines.}
\end{figure*}

To further elucidate the epitaxial quality of our ZnO films and to extract the in-plane and out-of-plane lattice constants, we recorded reciprocal space maps (RSM) for ZnO films with and without buffer layer. The result is shown in Figure~\ref{fig:rsm}. The observed elliptical shape of the $\mathrm{ZnO}(\overline{1}014)$ reflections stems from the difference in the FWHM of out-of-plane and in-plane mosaic spread. From their positions, we extract the same lattice parameters $a=0.3250$\,nm and $c=0.5205$\,nm for both a 300\,nm thin ZnO film without buffer layer and the 120\,nm thin ZnO film with a 15\,nm thin ZnO buffer layer from Figure~\ref{fig:process-buffer}. These values are very close to ZnO bulk ($a_0=0.32501$\,nm, $c_0= 0.52042$\,nm), attesting a relaxed growth of our ZnO thin films on Al$_2$O$_3$.

\subsubsection{Microscopic characterization.} \label{sec:tem}

\begin{figure*}
    \centering{\includegraphics[width=0.95\textwidth]{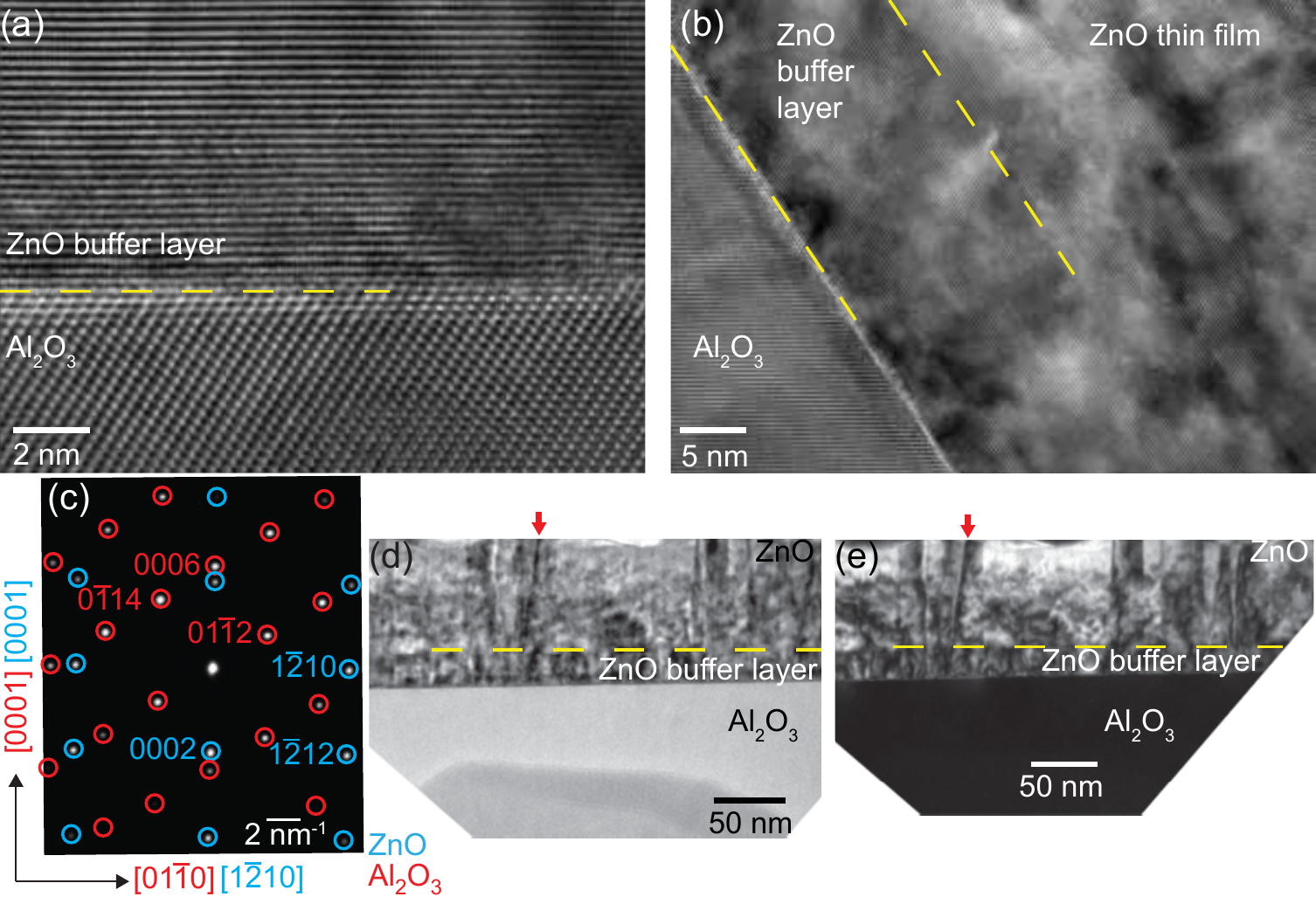}}
    \caption{\label{fig:tem}
             Transmission electron microscopy (TEM) analysis of a 120\,nm thin ZnO film on a 15\,nm thin ZnO buffer layer on a (0001)-oriented Al$_2$O$_3$ substrate.
             (a) Cross-sectional high-resolution TEM image of the interface region between Al$_2$O$_3$ substrate and ZnO buffer layer. It shows an abrupt and smooth interface, marked by the horizontal dashed line.
             (b) Large area cross-sectional TEM image with the interfaces indicated by dashed lines. The interface between the ZnO buffer layer and the ZnO thin film is not clearly visible.
             (c) TEM diffraction image. The diffraction peaks of the Al$_2$O$_3$ substrate (red) and ZnO (blue) are marked and labeled. The diffraction image confirms the epitaxial relations from XRD: $\mathrm{Al}_2\mathrm{O}_3[01\overline{1}0]\|\mathrm{ZnO}[1\overline{2}10]$.
             (d) Bright field and
             (e) dark field TEM micrograph, making the presence of dislocations and stacking faults visible. The red arrows indicate the same sample position.}
\end{figure*}

The high structural quality of our epitaxial ZnO thin film samples with buffer layer is further confirmed by the high-resolution transmission electron microscopy (HR-TEM) images shown in Figure~\ref{fig:tem}. The images were obtained from the very same sample described above (cf. Figs.~\ref{fig:process-buffer} and \ref{fig:rsm}). The HR-TEM micrograph of the 15\,nm thin ZnO buffer layer shown in Figure~\ref{fig:tem}(a) displays a sharp interface to the Al$_2$O$_3$ substrate. On a larger scale, the interface between the ZnO buffer layer and the ZnO thin film cannot be clearly discerned [Figure~\ref{fig:tem}(b)]. The TEM diffraction pattern shown in Figure~\ref{fig:tem}(c) confirms our HR-XRD results. ZnO grows relaxed on Al$_2$O$_3$ with the epitaxial relationship $\mathrm{ZnO}[1\overline{2}10]\|\mathrm{Al}_2\mathrm{O}_3[01\overline{1}0]$ sketched in Figure~\ref{fig:ZnO-Al2O3}.

Bright field and dark field micrographs (Figure~\ref{fig:tem}(d,e)) were obtained to investigate the density of dislocations. For the ZnO buffer layer, we estimate a dislocation density of the order of $10^{13}$\,cm$^{-2}$. Interestingly, the density of dislocations is not significantly lower in the ZnO thin film on top of the buffer. Moreover, the ZnO thin film shows no correlation of dislocations from the buffer and does even form new ones. The obtained bright field and dark field images show the same ZnO quality as the ones presented in Ref.~\cite{Petukhov2011} for a buffer-free ZnO thin film on Al$_2$O$_3$.

\subsection{Operation of the atomic radical source} \label{sec:atom-source}

\begin{figure*}[tb]
    \centering{\includegraphics[width=0.95\textwidth]{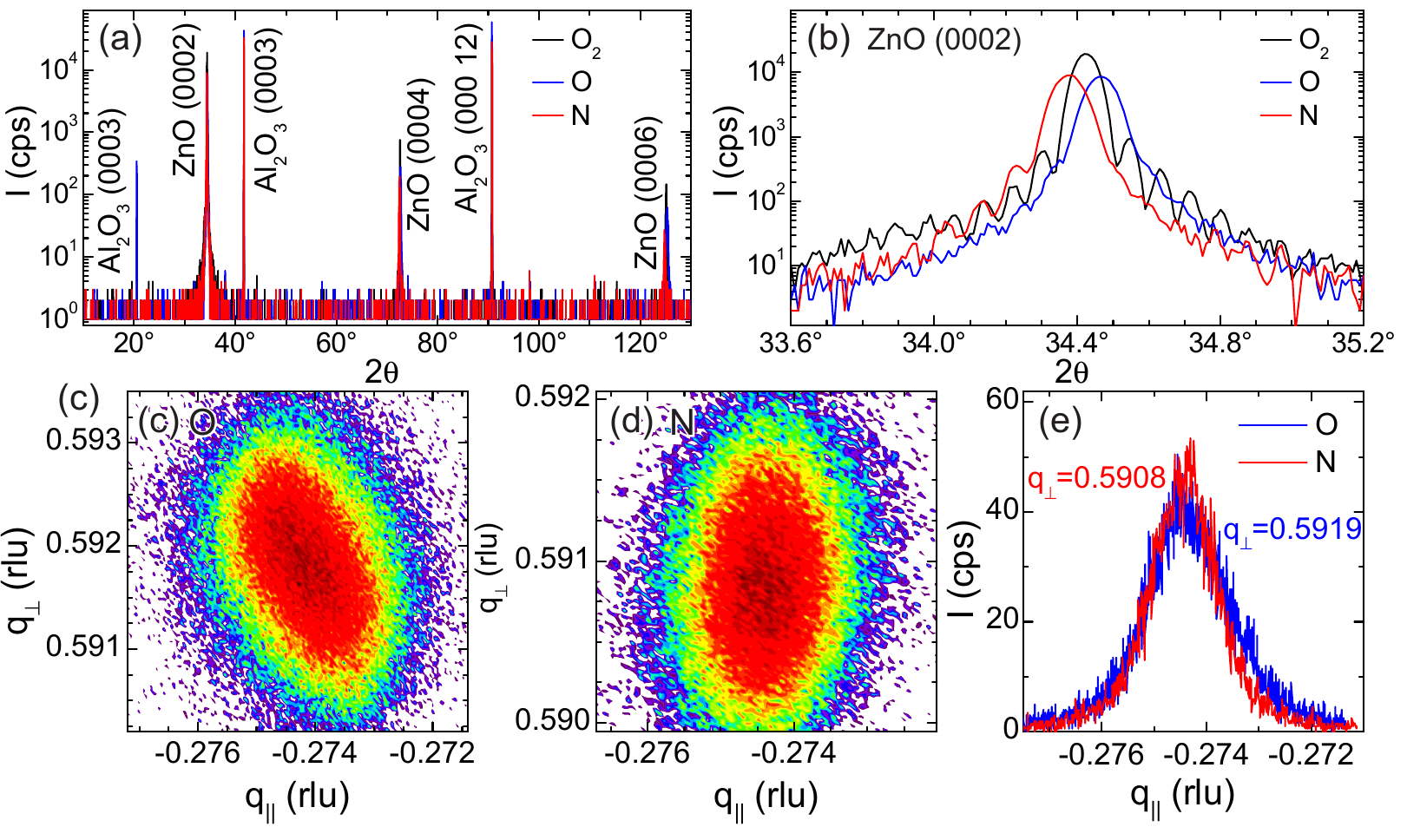}}
    \caption{\label{fig:atom-source}
              HR-XRD analysis of ZnO films grown in atomic N (red), atomic O (blue), and molecular O$_2$ (black) atmospheres. Deposition parameters: 2000 laser pulses, $\rho_\mathrm{L}=1$~J/cm$^2$, $p = 10^{-3}$\,mbar, $T_\mathrm{sub}=400^\circ$C.
             (a,b) Out-of-plane $\omega$-$2\theta$ diffraction scans.
             (c) Reciprocal space map (RSM) around the ZnO $(\overline{1}014)$ reflection from the film grown in O.
             (d) RSM around the ZnO $(\overline{1}014)$ reflection from the film grown in N.
             (e) $q_\|$-scans from the films of (c) and (d).}
\end{figure*}

To further optimize the ZnO thin film quality we studied the influence of an atomic background atmosphere, supplied by a radical source operated during deposition. We fabricated two ZnO thin film samples with buffer layer, one in atomic O and one in atomic N atmosphere at $10^{-3}$~mbar, while keeping otherwise optimum parameters ($\rho_\mathrm{L}=1$~J/cm$^2$, $T_\mathrm{sub}=400^\circ$C). We compare these two films to a reference sample grown in molecular O$_2$ atmosphere as described in the previous subsections. The experimental results of the structural characterization via HR-XRD are compiled in Figure~\ref{fig:atom-source}.

As shown by Figure~\ref{fig:atom-source}(a), the out-of-plane $\omega$-$2\theta$-scans do not reveal the formation of any secondary phases, neither when exposing the sample to atomic O nor to atomic N. On an enlarged scale around the ZnO(0002) reflection shown in Figure~\ref{fig:atom-source}(b), however, there are some differences visible between the films. The position of the ZnO(0002) reflection shifts to lower angles for films grown in atomic N (indicating a larger $c$-axis parameter of the unit cell) or to higher angles for films grown in atomic O (indicating a smaller $c$-axis parameter) compared to the reference thin film. To get further information and to allow for the determination of the in-plane and out-of-plane lattice strains, we recorded RSM for the films grown in atomic O and N, respectively (Figure~\ref{fig:atom-source}(c,d)). The peak positions of the $\mathrm{ZnO}(\overline{1}014)$ reflections appear at nearly the same in-plane reciprocal lattice vector $q_\|=-0.274\,\mathrm{rlu}$, whereas their out-of-plane components display a significant deviation. We observe $q_\bot=0.592\,\mathrm{rlu}$ for the film grown in atomic O and a slightly smaller position of $q_\bot=0.591\,\mathrm{rlu}$ for the one grown with atomic N. This corresponds to a larger $c$-axis lattice parameter for the latter which has already been indicated by the $\omega$-$2\theta$-scan [cf. Figure~\ref{fig:atom-source}(b)]. Its in-plane position, however, seems to be independent of the atomic species. This becomes more obvious from $q_\|$-linescans for both samples across the maxima of their respective $\mathrm{ZnO}(\overline{1}014)$ reflections [Figure~\ref{fig:atom-source}(e)]. The linescans display peaks at identical positions of $q_\|=-0.2745\,\mathrm{rlu}$, providing clear evidence for the same in-plane strain state for both samples.

In summary, the strain state of (0001)-oriented ZnO thin film samples on (0001)-oriented Al$_2$O$_3$ substrates does only weakly depend on the operation of an additional atom radical source during deposition. While the in-plane lattice constants largely remain unaffected, the $c$-axis lattice parameter decreases when growing in atomic O atmosphere. This result could point to the existence of oxygen vacancies ($V_\mathrm{O}$) in our ZnO films which are filled with atomic oxygen. Moreover, the $c$-axis lattice parameter becomes larger when the film is grown in the presence of atomic N. Keeping in mind that the ionic radius for 4-fold coordination of N$^{3-}$ (146\,pm) is slightly larger than the one of O$^{2-}$ (138\,pm), we conclude that nitrogen is partially incorporated into the ZnO lattice. This result is consistent with reports of nitrogen incorporation, e.g.~as molecular N$_2$ on O-sites (split interstitials) \cite{Schauries2013} or the association of N with a defect proton as $\mathrm{NH}^\times_0$ \cite{Jokela2007}. More information about the defect chemistry of N in oxides can be found in Ref.~\cite{Polfus2013}. We note that N incorporation does not result in an expansion of the unit cell within the film plane since the ZnO lattice is somehow clamped to the rigid Al$_2$O$_3$ substrate, but affects the out-of-plane lattice parameter.

\section{Deposition of ZnO-based heterostructures} \label{sec:heterostructures}

For semiconductor spintronic devices in general and for electrical spin injection experiments in particular, the fabrication of multilayer heterostructures is mandatory. We have shown recently that the oxide semiconductor ZnO can be combined with the ferromagnetic half-metallic oxide Fe$_3$O$_4$ \cite{Boger2008}. Here, we focus on the technical details of the deposition of a vertical spin valve device, based on a Co/ZnO/Ni trilayer system \cite{Althammer2012}. We successfully applied this device for the electrical investigation of the transport and the dephasing properties of spin-polarized charge carriers in ZnO \cite{Althammer2012}.

\subsection{Growth parameters}

The multilayer system is built up (from bottom to top) of a 12\,nm thick metallic TiN layer as bottom electrode, a 11\,nm thick Co layer as the first ferromagnetic electrode, a semiconducting ZnO layer with variable thickness from 15\,nm to 100\,nm, a 11\,nm thick Ni layer as the second ferromagnetic electrode, and finally a 24\,nm thick metallic Au layer as top electrode. TiN and ZnO were deposited in the laser-MBE chamber, Co, Ni, and Au in a metallization chamber attached to the same UHV cluster system. In this way, all layers can be grown on top of each other without breaking vacuum.

\begin{figure*}[tb]
    \centering{\includegraphics[width=0.95\textwidth]{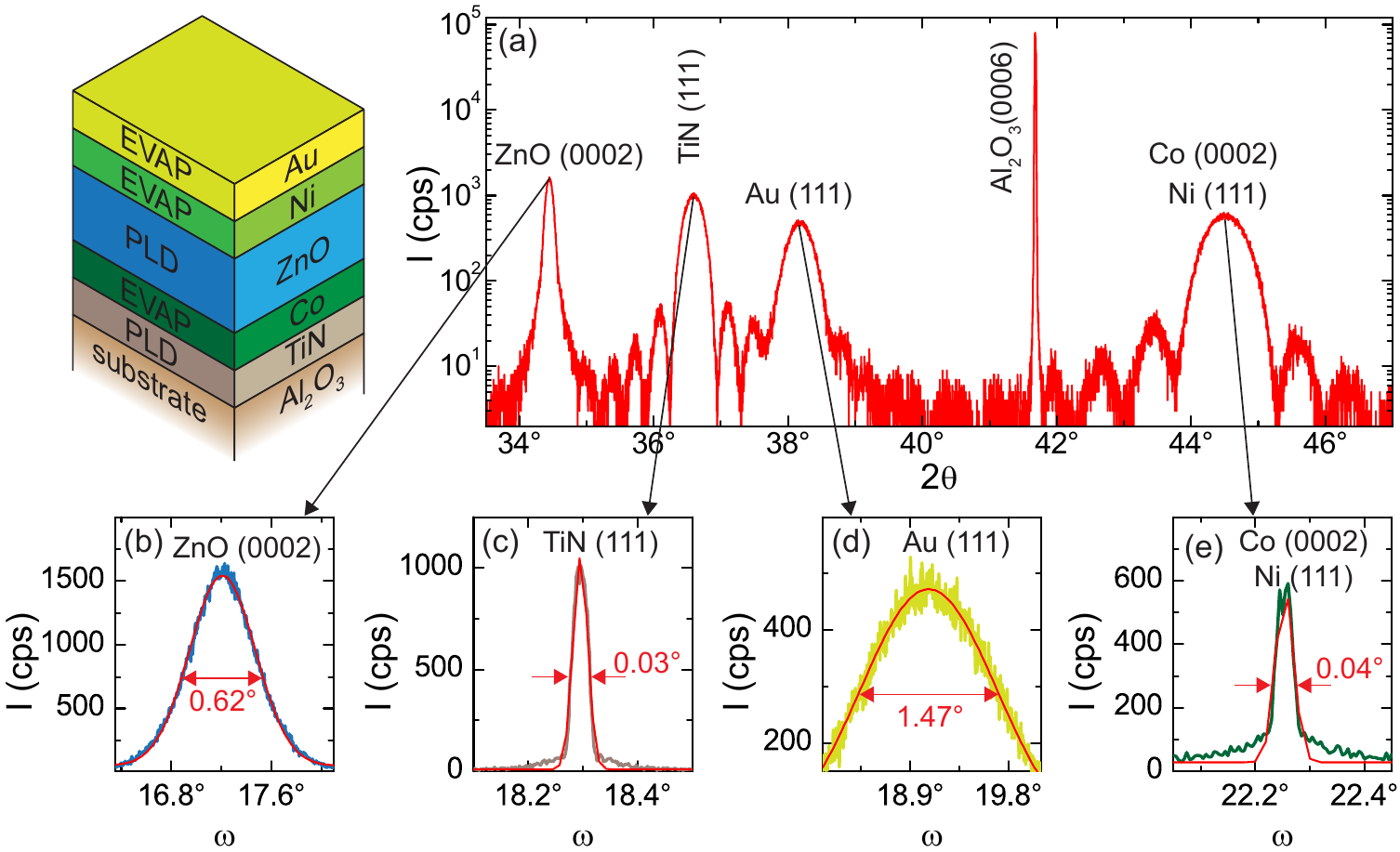}}
    \caption{\label{fig:heterostructure}
             HR-XRD results from a TiN(12~nm)/ Co(11~nm)/ ZnO(80~nm)/ Ni(11~nm)/ Au(24~nm) multilayer stack, deposited by pulsed laser deposition (PLD) or electron beam evaporation (EVAP) on a (0001)-oriented Al$_2$O$_3$ substrate.
             (a) Out-of-plane $\omega$-$2\theta$ diffraction scan on an enlarged scale around the Al$_2$O$_3$(0006) reflection. No secondary phases are visible. Satellites due to Laue oscillations are clearly resolved for TiN(111) and Co(0002).
             (b-e) Out-of-plane $\omega$ rocking curves around the (b) ZnO(0002), (c) TiN(111), (d) Au(111), and (e) Co(0002)/Ni(111) reflections. The red lines represent Gaussian fits to the data, the FWHM values are given and indicated by arrows.}
\end{figure*}

TiN is grown on the Al$_2$O$_3$ substrate via pulsed laser deposition from a polycrystalline, stoichiometric target with a repetition rate of 10\,Hz and a laser fluence of $\rho_{L} = 2$J/cm$^2$ at a temperature of $T_\mathrm{sub}=600^\circ$C in an inert atmosphere of Ar at $7\,\times\,10^{-4}$\,mbar \cite{Reisinger2003b}. The sample is cooled down to room temperature and the process chamber is evacuated to its base pressure. The sample is then transferred \textit{in-situ} to the metallization chamber. There, the Co layer is deposited at room temperature at $p<10^{-7}$\,mbar via electron beam evaporation of metallic Co. The sample is then again transferred back to the laser-MBE chamber where it is heated to $T_\mathrm{sub}=400^\circ$C in an Ar atmosphere of $10^{-3}$\,mbar to prevent oxidation of the ferromagnetic Co electrode. The semiconducting ZnO layer is grown via pulsed laser deposition as described in section~\ref{sec:ZnO thin films}, but with a repetition rate of 10\,Hz and $\rho_\mathrm{L}=1$\,J/cm$^2$ at $T_\mathrm{sub}=400^\circ$C. Note that we keep the inert Ar atmosphere during ZnO deposition to prevent oxidation of the Co layer. The sample is cooled down to room temperature, the process chamber is evacuated to base pressure, and the sample transferred again \textit{in-situ} to the metallization chamber. Finally, the Ni and Au layers are deposited at room temperature and $p<10^{-7}$\,mbar via electron beam evaporation. The resulting multilayer stack is illustrated in Figure~\ref{fig:heterostructure}.

\subsection{Structural characterization}

The structural quality of the multilayer stack was investigated by HR-XRD as described in section~\ref{sec:ZnO thin films}. The out-of-plane $\omega$-$2\theta$-scan shown in Figure~\ref{fig:heterostructure}(a) does not reveal any secondary phases and demonstrates the high structural quality of the sample. All detected reflections can be assigned to the respective materials of the spin valve multilayer structure. The cubic materials TiN, Ni, and Au grow (111)-oriented, the hexagonal Co and ZnO layers (0001)-oriented. The signals from Co and Ni can hardly be separated as the $c$-axis lattice parameter of hcp Co ($c_\mathrm{Co}=0.407$\,nm \cite{Martienssen2005}) is close to twice the lattice spacing $d_{111} = \frac{1}{3} \sqrt{3} a = 0.203$\,nm of the $[111]$-planes in cubic Ni with $a_{\mathrm Ni} = 0.352$\,nm \cite{Martienssen2005}. Satellites around the TiN(111) reflection ($2\theta=36.60^\circ$), the Au(111) reflection ($38.17^\circ$), and the Co(0002)/Ni(111) reflection ($44.50^\circ$) due to Laue oscillations are clearly visible, demonstrating the coherent growth with small interface roughness and indicating a high structural quality. From the out-of-plane reflections, we calculate the corresponding lattice parameters for each layer ($a_\mathrm{TiN}=0.425$\,nm, $c_\mathrm{Co}=0.407$\,nm, $c_\mathrm{ZnO}=0.521$\,nm, $a_\mathrm{Ni}=0.352$\,nm, $a_\mathrm{Au}=0.408$\,nm) and find them very close ($<0.1\%$) to the respective bulk values \cite{Martienssen2005}. This indicates a relaxed growth of the epilayers on each other and an unstrained multilayer system.

To obtain further information on the structural quality we studied the $\omega$ rocking curves for each layer. As shown by Figure~\ref{fig:heterostructure}(b), the rocking curve of the ZnO(0002) reflection is relatively broad with a FWHM of $0.62^\circ$. But one should keep in mind that the ZnO has to grow on top of the Co layer with a very large in-plane lattice mismatch ($>25\%$), making the achieved value still exceptionally good. For the rocking curve of the TiN(111) reflection shown in Figure~\ref{fig:heterostructure}(c), a FWHM of $0.03^\circ$ is achieved, demonstrating a very high structural quality of this bottom electrode layer. For the rocking curve of the Au(111) reflection [Figure~\ref{fig:heterostructure}(d)] we obtain a large FWHM of $1.47^\circ$. Finally, as shown in Figure~\ref{fig:heterostructure}(e), for the rocking curve around the Co(0002)/Ni(111) reflection, we obtain a FWHM of $0.04^\circ$, attesting a low mosaic spread of the Co layer. From this analysis and not surprising, we see that the FWHM of the rocking curves increases with each successive layer deposited in the multilayer structure. Nevertheless, the obtained values point to an overall low mosaic spread of the respective epitaxial layers. With regard to ZnO, its structural quality in the multilayer system is lower than before for the ZnO single layers (see section~\ref{sec:ZnO thin films}). There, we reported a FWHM of $0.03^\circ$ for the ZnO(0002) reflection in our optimized thin films. We attribute this decrease in ZnO quality to the TiN/Co template and the use of the Ar atmosphere during ZnO growth.

\begin{figure}[tb]
    \centering{\includegraphics[width=0.5\textwidth]{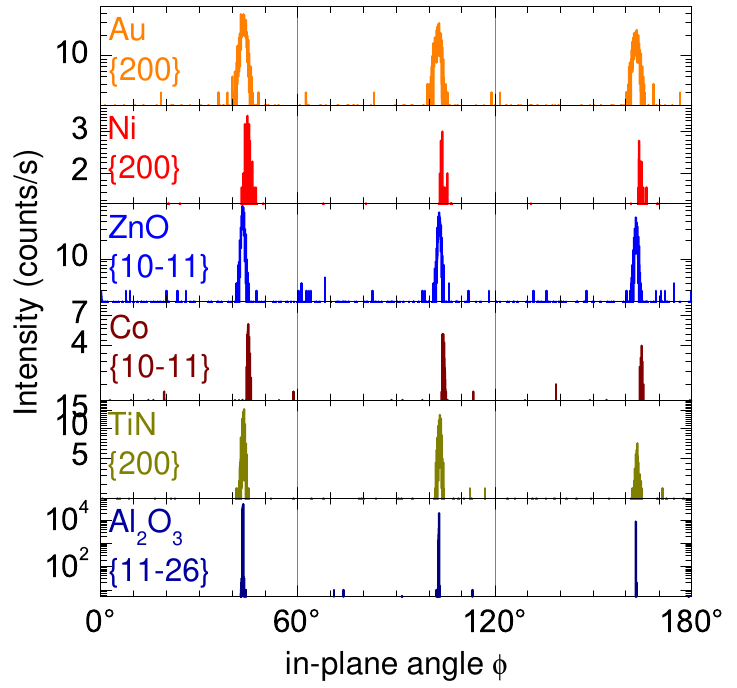}}
    \caption{\label{fig:phi-scan}
             XRD $\phi$-scans from the same multilayer stack as in Fig.~\ref{fig:heterostructure}. The six data sets correspond to the (a) Au$(200)$, (b) Ni$(200)$, (c) ZnO$(10\overline{1}1)$, (d) Co$(10\overline{1}1)$, (e) TiN$(200)$ thin film reflections, and the (f) Al$_2$O$_3(11\overline{2}6)$ substrate reflection. The diffraction peaks line up at identical $\phi$ values, indicating a well formed epitaxial in-plane orientation of all deposited layers.}
\end{figure}

To investigate the in-plane epitaxial relationship between the layers, we performed $\phi$-scans for asymmetric thin film and substrate reflections. The $\phi$-scans were carried out for each reflection and exhibit a 6-fold symmetry for all layers. As shown in Figure~\ref{fig:phi-scan}, the $\phi$-scans for each layer display six distinctive peaks at the same $\phi$ angles, namely at $43^\circ, 103^\circ, 163^\circ, 223^\circ, 283^\circ, 343^\circ$. Form this we derive the following in-plane epitaxial relationships: $\mathrm{Au}(111)[2\overline{1}\overline{1}] \|$ $\mathrm{Ni}(111)[2\overline{1}\overline{1}] \|$ $\mathrm{ZnO}(0001)[10\overline{1}0] \|$ $\mathrm{Co}(0001)[10\overline{1}0] \|$ $\mathrm{TiN}(111)[2\overline{1}\overline{1}] \|$ $\mathrm{Al}_2\mathrm{O}_3(0001)[11\overline{2}0]$. As the selected $\mathrm{Al}_2\mathrm{O}_3\{11\overline{2}6\}$ reflection directly represents the orientation of the close-packed O direction, it becomes evident that each layer is aligned to this sublattice. This has already been reported for TiN(111) \cite{Talyansky1999}, Co(0001) \cite{Ago2010}, ZnO(0001) \cite{Chen1998}, and Au(111) thin films \cite{Kastle2002} all directly grown on (0001)-oriented Al$_2$O$_3$. Our results prove that the respective in-plane orientations of the single layers are preserved when grown on top of each other.

\section{Summary} \label{sec:summary}

We have grown high quality epitaxial ZnO thin films and ZnO based heterostructures on sapphire substrates by laser molecular beam epitaxy (MBE). The growth was realized with an advanced laser-MBE system, making use of several recent developments. We particularly discuss the use of a flexible ultra-violet laser beam optics allowing for the precise variation of the laser fluence, an infrared laser heating system allowing for both high substrate temperatures in UHV environment and rapid temperature changes, and the use of a source for atomic oxygen and nitrogen.

As an application of our advanced laser-MBE system the growth of epitaxial ZnO thin films and ZnO based heterostructures is presented. We describe the systematic optimization of the deposition parameters for single ZnO films such as laser fluence and substrate temperature as well as the use of ZnO buffer layers. The film growth was monitored by \textit{in situ} RHEED. The detailed structural characterization by x-ray analysis and transmission electron microscopy shows that epitaxial ZnO thin films with high structural quality can be achieved. For films grown without any buffer layer, the rocking curves of the ZnO(0002) reflection show a FWHM as small as $0.04^\circ$, indicating a very small out-of-plane mosaic spread. For the rocking curve of the asymmetric ZnO$(10\overline{1}1)$ reflection, a FWHM of $0.4^\circ$ was obtained, demonstrating that also the in-plane mosaic spread is small. For the in-plane orientation we found $\mathrm{ZnO}[10\overline{1}0]\|\mathrm{Al}_2\mathrm{O}_3[11\overline{2}0]$, confirming the in-plane parallel alignment of the O-sublattices of ZnO and Al$_2$O$_3$. Employing a ZnO buffer layer the out-of-plane mosaic spread of the ZnO film grown on this buffer could be slightly improved, whereas the in-plane mosaic spread stayed about the same and is determined by the buffer layer. The same is true for the dislocation density which is found to be of the order of $10^{13}$\,cm$^{-2}$. The use of atomic nitrogen results in a slight increase of the $c$-axis lattice parameter, suggesting the incorporation of a finite amount of nitrogen into the ZnO film.

We also realized the heteroepitaxial growth of ZnO based multilayers as a prerequisite for spin transport experiments and the fabrication of spintronic devices. As an example, we showed that TiN/Co/ZnO/Ni/Au multilayer stacks can be grown on (0001)-oriented sapphire with good structural quality of all layers and well defined in-plane epitaxial relations.

\ack
We acknowledge technical support by Andreas Erb in the preparation of the polycrystalline targets.
We thank Evgeniy Zamburg and Eva-Maria Karrer-M\"{u}ller for the fabrication and characterization of the thin film samples grown with use of the atomic radical source and the spin valve multilayer samples, respectively.
We are grateful to Sven-Martin H\"{u}hne in the group of Werner Mader at the Rheinische Friedrich-Wilhelms-Universit\"{a}t in Bonn, Germany, for the HR-TEM analysis.
Finally, we acknowledge financial support by the Deutsche Forschungsgemeinschaft via SPP 1285 (Project No.~GR 1132/14) and the German Excellence Initiative via the ``Nanosystems Initiative Munich (NIM)''.

\section*{References}

\bibliographystyle{unsrt}
\bibliography{Opel_PLD_ZnO_05July2013}


\end{document}